\documentclass[a4paper,11pt]{article}
\usepackage[dvips]{graphicx}
\usepackage{amsmath}
\usepackage{amssymb}
\usepackage{cite}
\usepackage{cancel}
\usepackage{fancybox}
\usepackage[dvips]{color}
\usepackage{ulem}
\usepackage{pifont}
\usepackage{colortbl}
\usepackage{multicol}

\usepackage{colortbl}
\usepackage{multicol}
\usepackage[all,knot]{xy}
\usepackage{longtable}

\makeatletter
 
 \@addtoreset{equation}{section}
\makeatother

\newcommand{\utxtarrow}[1]{\xrightarrow[]{\hspace{1mm}#1\hspace{1mm}}}

\usepackage{xcolor}

\begin{document}

\begin{titlepage}

\begin{flushright}
 KYUSHU-HET-272\\
\end{flushright}

\begin{center}

\vspace{0cm}
{\Large\textbf{
Pseudo-Nambu-Goldstone Dark Matter\\
 in $SU(7)$ Grand Unification
}}
\vspace{1cm}

\renewcommand{\thefootnote}{\fnsymbol{footnote}}
Cheng-Wei Chiang${}^{1,2}$\footnote[2]{chengwei@phys.ntu.edu.tw},
Koji Tsumura${}^{3}$\footnote[3]{tsumura.koji@phys.kyushu-u.ac.jp},
Yoshiki Uchida${}^{4,5}$\footnote[4]{uchida.yoshiki@m.scnu.edu.cn},
and Naoki Yamatsu${}^{1}$\footnote[1]{yamatsu@phys.ntu.edu.tw}
\vspace{5mm}

\textit{
 $^1${Department of Physics and Center for Theoretical Physics,\\
 National Taiwan University, Taipei, Taiwan 10617, R.O.C.}\\
 $^2${Physics Division, National Center for Theoretical Sciences,\\
 Taipei, Taiwan 10617, R.O.C.}\\
 $^3${Department of Physics, Kyushu University,\\
 744 Motooka, Nishi-ku, Fukuoka, 819-0395, Japan}\\
 $^4${Key Laboratory of Atomic and Subatomic Structure and Quantum Control (MOE),\\
 Guangdong Basic Research Center of Excellence for Structure and Fundamental Interactions of Matter, Institute of Quantum Matter,\\ 
 South China Normal University, Guangzhou 510006, China}\\
 $^5${Guangdong-Hong Kong Joint Laboratory of Quantum Matter, Guangdong Provincial Key Laboratory of Nuclear Science, Southern Nuclear Science Computing Center,\\ 
 South China Normal University, Guangzhou 510006, China}
}

\date{\today}

\abstract{
We propose a grand unified theory (GUT) pseudo-Nambu-Goldstone boson (pNGB) dark matter (DM) model based on $SU(7)$ gauge symmetry.  In the GUT model, the Standard Model (SM) gauge symmetry $G_{\rm SM} := SU(3)_C\times SU(2)_L\times U(1)_Y$ and the  ``dark'' gauge symmetry $SU(2)_D$ are unified, where the $SU(2)_D$ symmetry plays an important role in the stability of DM.  The unification of SM fermions and dark sector fermions is partially realized.  The gauge symmetry $SU(7)$ is spontaneously broken to $SU(5)\times SU(2)\times U(1)$ gauge symmetry at the GUT scale by the nonvanishing vacuum expectation values of an $SU(7)$ adjoint scalar field, where the $SU(5)$ gauge symmetry is not usual $SU(5)$ GUT gauge symmetry.  The symmetry is further broken to $G_{\rm SM}\times SU(2)_D$ at an intermediate scale.  Furthermore, the $SU(2)_D$ symmetry is broken by the $SU(2)_D$ doublet and triplet scalar fields at the TeV scale.  In the pNGB DM model based on $G_{\rm SM}\times SU(2)_D$, the residual global $U(1)_V$ dark custodial symmetry guarantees DM stability. On the other hand, in the $SU(7)$ pNGB DM model, this global symmetry is explicitly broken by the Yukawa interaction and the effective Majorana mass terms.  In the scalar sector, the cubic coupling constants of the $SU(2)_D$ doublet and triplet scalar fields are the order parameters of the $U(1)_V$ symmetry breaking.  To maintain $U(1)_V$ symmetry and thus the DM stability, we need to tune Yukawa coupling constants and cubic scalar couplings at high accuracy.  We find that the allowed DM mass region is quite restricted as the gauge coupling constant of $SU(2)_D$ is determined by the condition of the gauge coupling unification.  To satisfy gauge coupling unification and the current experimental constraint on proton lifetime, we find that three generations of $SU(3)_C$ adjoint fermions and another three generations of $SU(2)_L$ adjoint fermions with the intermediate mass scale are required.  We also find that there is no other solution to satisfy simultaneously the gauge coupling unification and the proton decay constraint if one assumes the other symmetry breaking schemes. 
}

\end{center}
\end{titlepage}



\section{Introduction}

One of the most attractive ideas to construct theories beyond
the standard model (SM) is the grand unification of gauge interactions \cite{Georgi:1974sy}. 
As is well-known \cite{Slansky:1981yr,Yamatsu:2015gut}, 
the only candidates for grand unified theory (GUT) gauge group 
in four spacetime dimensional theories
are $SU(n) (n\geq 5)$\cite{Georgi:1974sy,Inoue:1977qd},
$SO(4n+2) (n\geq 2)$\cite{Fritzsch:1974nn,Ida:1980ea,Fujimoto:1981bv}, and
$E_6$\cite{Gursey:1975ki} because of the appropriate ranks and types of representations.
(In higher dimensional theories, not only these groups 
\cite{Kawamura:1999nj,Hall:2001pg,Lim:2007jv,Kojima:2011ad,Yamatsu:2017sgu,Yamatsu:2018fsg}
but also other gauge groups 
\cite{Nomura:2008sx,Hosotani:2015hoa,Yamatsu:2017ssg}
are available.)
The GUT models predict the unification of the SM gauge coupling
constants at the GUT scale because the SM gauge symmetries
$SU(3)_C$, $SU(2)_L$, and $U(1)_Y$, unrelated within the SM, is
unified into one GUT gauge symmetry.
In the GUT framework, the SM gauge fields associated with gauge symmetries
are unified into a single GUT gauge field and, furthermore, quarks and
leptons belonging to different representations under the SM gauge
symmetries are unified into the same fermion multiplets. 
As a result, such unification of quarks and leptons leads to processes with baryon 
and lepton number violations, and thus predicts the very
interesting phenomena of proton decay.
So far, no proton decay has been discovered yet, but experiments
are under way\cite{Super-Kamiokande:2017gev,Super-Kamiokande:2020tor,Super-Kamiokande:2020wjk,Super-Kamiokande:2022egr,JUNO:2022qgr} 
and will be upgraded in the near future\cite{Hyper-Kamiokande:2018ofw}.

As has been discussed in various GUT models, the unification scale for
the gauge coupling constants is expected to be of
$\mathcal{O}(10^{15}-10^{18})$~GeV.
This scale must satisfy experimental constraints from the searches of proton
decay mediated by leptoquark gauge bosons \cite{Heeck:2019kgr}, and
is usually expected to be below the Planck scale.
In addition, the existence of an intermediate scale between the
electroweak and the grand unification scales is helpful in achieving gauge
coupling unification in the framework of non-supersymmetic GUT models.
Also, the tiny neutrino masses from the neutrino oscillation data seem
to suggest an intermediate scale of $\mathcal{O}(10^{10}-10^{14})$~GeV
through the so-called seesaw mechanism \cite{Minkowski:1977sc,Yanagida:1979as}.
Therefore, the intermediate scale required to realize the unification of
gauge coupling constants may naturally explain the tiny neutrino masses
by identifying it with the ``right-handed'' neutrino scale.

The SM explains the results of accelerator experiments with the
exception of only a few anomalous results. 
From non-accelerator experiments, however, phenomena beyond the SM expectations emerge.
The existence of dark matter (DM) has already been confirmed by several
astronomical observations such as spiral galaxies
\cite{Corbelli:1999af,Sofue:2000jx}, gravitational lensing 
\cite{Massey:2010hh}, cosmic microwave background radiation \cite{Aghanim:2018eyx},
and the collision of the Bullet Cluster \cite{Randall:2007ph}.
However, there is no particle in the SM that can serve as a candidate for the DM.
It thus becomes one of the key issues in modern particle physics and cosmology to
pinpoint the origin and identity of the DM.

Since the nature of DM is not yet fully understood, there exist many candidate models. One such candidate is called a weakly interacting massive particle (WIMP)\cite{Arcadi:2017kky}, with attractive properties that the DM particles are produced thermally by non-gravitational interactions and that its relic abundance is fixed automatically by the freeze-out mechanism. 
To achieve the correct relic abundance, the mass of the WIMP is expected to be in
the range of $\mathcal{O}(10)$~GeV to $\mathcal{O}(100)$~TeV. Many DM direct 
and indirect detection experiments are designed and running to probe its existence through its non-gravitational interactions. However, no clear evidence of WIMP has been observed yet, and the direct detection 
experiments have yielded strong constraints on WIMP mass and interactions.
To avoid the severe constraints of DM direct detection experiments,
several models have been considered in the WIMP DM scenario, such as a fermionic DM with pseudo-scalar interactions
\cite{Freytsis:2010ne,Ipek:2014gua,Arcadi:2017wqi,Sanderson:2018lmj,Abe:2018emu,Abe:2019wjw}
and a pseudo-Nambu-Goldstone boson (pNGB) DM
with additional $U(1)$ symmetry \cite{Barger:2010yn,Barducci:2016fue,Gross:2017dan,Balkin:2017aep,Ishiwata:2018sdi,Huitu:2018gbc,Cline:2019okt,Jiang:2019soj,Arina:2019tib,Karamitros:2019ewv,Abe:2020iph,Okada:2020zxo,Zhang:2021alu,Abe:2021jcz,Abe:2021vat,Zeng:2021moz}
or $SU(2)$ symmetry \cite{Abe:2022mlc,Otsuka:2022zdy}
in the non-GUT framework or in the GUT framework
\cite{Abe:2021byq,Okada:2021qmi}. 

As pointed out in the original pNGB DM model \cite{Gross:2017dan}, the DM
has the property of a Nambu-Goldstone (NG) mode such that the coupling of the DM with the SM Higgs boson is proportional to its momentum. As a result, the scattering cross sections of the DM with the SM particles via the exchange of the Higgs boson are strongly suppressed by the small momentum transfer, while the annihilation cross section associated with the freeze-out mechanism is maintained at an appropriate level.

Recently, a new pNGB DM model based on the SM gauge
symmetry and a non-Abelian dark gauge symmetry $SU(2)_D$ has been proposed \cite{Otsuka:2022zdy},
in which scalars in the ${\bf 2}$ and ${\bf 3}$ representations of $SU(2)_D$ are introduced.
The $SU(2)_D$ gauge symmetry is spontaneously broken to the exact
$U(1)_V$ global symmetry by the vacuum expectation values (VEVs) of the ${\bf 2}$ and ${\bf 3}$ 
scalars. 
The resulting pNGB is charged under the $U(1)_V$ custodial symmetry and identified as the DM. 
The exact $U(1)_V$ symmetry is a remnant of the enlarged global symmetry of $SU(2)_D$ scalar sector, 
which guarantees the stability of the DM.
The DM relic abundance can be correctly reproduced while escaping the severe
constraints from the direct detection experiments.

The purpose of this paper is to propose a GUT pNGB DM model based on
$SU(7)$ gauge symmetry that includes the SM gauge symmetry 
$G_{\rm SM}$ and the dark gauge symmetry $SU(2)_D$.
The low-energy effective theory of the GUT model is almost the same as 
the pNGB DM model based on $G_{\rm SM}\times SU(2)_D$ symmetry
discussed in Ref.~\cite{Otsuka:2022zdy}.
The unification of SM fermions and dark sector fermions is partially
realized due to $SU(7)$ gauge symmetry. 
Additional vectorlike fermions appear at the $SU(2)_D$ symmetry breaking scale. 
Although there are several symmetry-breaking schemes that can lead to
the subgroup symmetry $G_{\rm SM}\times SU(2)_D$, 
we find that the gauge coupling unification and the proton decay
constraints can be satisfied only by the symmetry-breaking pattern 
$SU(7)\to SU(5)_{CD}\times SU(2)_L\times U(1)_\alpha$,
where $SU(5)_{CD}$ encompasses $SU(3)_C$ and
$SU(2)_D$ rather than the usual $SU(5)$ grand unified group and the symmetry is spontaneously broken at the GUT
scale via the nonvanishing VEV of an $SU(7)$ adjoint scalar field. The symmetry is further broken to $G_{\rm SM}\times SU(2)_D$ at an intermediate scale. Furthermore, the $SU(2)_D$ symmetry is broken by the
VEVs of the $SU(2)_D$ doublet and triplet scalar fields at the TeV scale.
In the pNGB DM model based on $G_{\rm SM}\times SU(2)_D$, the $U(1)_V$
custodial symmetry guarantees DM stability. On the other hand, in the $SU(7)$ pNGB DM model, this global symmetry is explicitly broken by the Yukawa interaction and the effective Majorana mass terms.
In the scalar sector, the cubic coupling constants of the $SU(2)_D$ doublet and triplet scalar fields are the order parameters of the $U(1)_V$ symmetry breaking.
In this model, the $U(1)_V$ symmetry is essential for DM stability, and we need to tune the Yukawa coupling constants and cubic scalar couplings to ensure the symmetry at high accuracy.
We find that the allowed DM mass region is more limited than in the previous pNGB DM model because the gauge coupling constant of $SU(2)_D$ is now determined by the condition of the gauge coupling unification.

The paper is organized as follows.
In Sec.~\ref{Sec:Model}, we introduce the pNGB DM model in the $SU(7)$ grand unification framework.
In Sec.~\ref{Sec:RGE}, we show how to realize the gauge coupling unification and satisfy current experimental constraints from proton decays.
In Sec.~\ref{Sec:DM}, we show under what
conditions sufficient DM stability is satisfied.
In Sec.~\ref{Sec:Predictions}, we show the prediction of the model.
Section~\ref{Sec:Summary} has further discussions and summarizes our findings.

\section{pNGB DM and grand unification}
\label{Sec:Model}

In this section, we first introduce the pNGB DM model based upon the $SU(7)$ grand unified gauge group and study the symmetry-breaking schemes.  Afterward, we present the low-energy effective theory of the model, which is necessary for the discussions of gauge coupling unification and DM stability. 

\subsection{The model}

The model consists of an $SU(7)$ gauge field denoted by ${A}_\mu$; fermions
in the ${\bf 21}$, ${\bf \overline{7}}$, and ${\bf 48}$ representations denoted by
$\Psi_{\bf 21}^{(m)}$ $(m=1,2,3)$,
$\Psi_{\bf \overline{7}}^{(n)}$ $(n=1,2,...,9)$,
$\Psi_{\bf 48}^{(m)}$ $(m=1,2,3)$, respectively;
complex scalar fields in the ${\bf 7}$, ${\bf 21}$, and ${\bf 35}$ representations denoted by 
$\Phi_{\bf 7}$, $\Phi_{\bf 21}$, and $\Phi_{\bf 35}$, respectively; and 
a real scalar field in the ${\bf 48}$ representation denoted by $\Phi_{\bf 48}$.
The $SU(7)$ gauge field contains the $G_{\rm SM}$ and $SU(2)_D$ gauge fields. 
One of the three {\bf 21} fermions and three of the nine ${\bf \overline{7}}$ fermions form one generation of quarks and leptons plus vectorlike fermions under $G_{\rm SM}$. 
The existence of the three ${\bf 48}$ fermions plays important roles for
gauge coupling unification and seesaw mechanism to be discussed later.
The ${\bf 7}$ scalar field includes a complex scalar in the ${\bf 2}$ representation of $SU(2)_D$.
The ${\bf 48}$ real scalar field includes a real scalar in the ${\bf 3}$ representation of $SU(2)_D$.
As alluded to earlier,
these ${\bf 2}$ and ${\bf 3}$ scalar fields in $SU(2)_D$ 
play a crucial role in breaking the $SU(2)_D$ spontaneously
through their VEVs.
In order to reproduce the observed quark and lepton masses, the ${\bf 35}$ complex scalar field should include the SM Higgs boson. 
Therefore, the SM Higgs boson must be a linear combination of the scalars in ${\bf 35}$ and
${\bf 7}$ of $SU(7)$.
The scalar fields $\Phi_{\bf 21}$ and $\Phi_{\bf 48}$ are responsible
for breaking the $SU(7)$ symmetry to $G_{\rm SM}\times SU(2)_D$.
The field content of the $SU(7)$ GUT model is summarized in 
Table~\ref{Tab:Matter_Content-GUT-SU7}.

\begin{table}[tbh]
\begin{center}
\begin{tabular}{|c|c||c|c|c||c|c|c|c|c|}\hline
 \rowcolor[gray]{0.8}
 &{${A}_\mu$}
 &{${\Psi}_{\bf 21}^{(m)}$}
 &{${\Psi}_{\bf \overline{7}}^{(n)}$}
 &{${\Psi}_{\bf 48}^{(m)}$}
 &${\Phi}_{\bf 7}$
 &${\Phi}_{\bf 21}$
 &${\Phi}_{\bf 35}$
 &${\Phi}_{\bf 48}$
 \\\hline
 $SU(7)$
 &{${\bf 48}$}
 &{${\bf 21}$}
 &${\bf \overline{7}}$
 &{${\bf 48}$}
 &${\bf 7}$
 &${\bf 21}$
 &${\bf 35}$
 &${\bf 48}$
\\\hline
 $SL(2,\mathbb{C})$
 &$(1/2,1/2)$&$(1/2,0)$&$(1/2,0)$&$(1/2,0)$
 &$(0,0)$&$(0,0)$&$(0,0)$&$(0,0)$
 \\\hline
\end{tabular}
 \caption{\small The field content in the $SU(7)$ model, where
 $m=1,2,3$ and $n=1,2,...,9$.
Note that 
a linear combination of $\Phi_{\bf 7}$ 
and $\Phi_{\bf 35}$ includes the SM Higgs boson $H$,
$\Phi_{\bf 7}$ includes the scalar field in ${\bf 2}$ of $SU(2)_D$, 
and 
$\Phi_{\bf 48}$ contains the scalar field in ${\bf 3}$ of $SU(2)_D$.
 }
\label{Tab:Matter_Content-GUT-SU7}
\end{center}  
\end{table}

The most general Lagrangian of the model that contains all $SU(7)$ 
invariant and renormalizable terms is given by
\begin{align}
 {\cal L}=&
  -\frac{1}{2}\mbox{tr}
 \left[{F}_{\mu\nu}{F}^{\mu\nu}\right]
\allowdisplaybreaks[1]\nonumber\\
 &+
 \sum_{{\bf y}={\bf 7},{\bf 21},{\bf 35}}
 \left({D}_\mu{\Phi}_{\bf y}\right)^\dag
 \left({D}^\mu{\Phi}_{\bf y}\right)
 +\frac{1}{2}
 \left({D}_\mu{\Phi}_{\bf 48}\right)^T
 \left({D}^\mu{\Phi}_{\bf 48}\right)
 -V\left(\left\{{\Phi}_{\bf x}\right\}\right)
\allowdisplaybreaks[1]\nonumber\\
 &+
 \sum_{m=1}^3
 \overline{{\Psi}_{\bf 21}^{(m)}}i
 \cancel{D}{\Psi}_{\bf 21}^{(m)}
 +\sum_{n=1}^9
 \overline{{\Psi}_{\bf \overline{7}}^{(n)}}i
 \cancel{D}{\Psi}_{\bf \overline{7}}^{(n)}
 +
 \sum_{m=1}^3
 \overline{{\Psi}_{\bf 48}^{(m)}}i
 \cancel{D}{\Psi}_{\bf 48}^{(m)}
\allowdisplaybreaks[1]\nonumber\\
 &-
 \sum_{m,n=1}^3 \frac12
 M_{M}^{(mn)}
 {\Psi}_{\bf 48}^{(m)}{\Psi}_{\bf 48}^{(n)}
\allowdisplaybreaks[1]\nonumber\\
 &+\bigg(-\sum_{m,n=1}^{3}
y_1^{(mn)}
{\Phi}_{\bf 35}
 \left({\Psi}_{\bf 21}^{(m)}{\Psi}_{\bf 21}^{(n)}\right)_{\overline{\bf 35}}
 -\sum_{m=1}^{3}\sum_{n=1}^{9}
y_2^{(mn)}
{\Phi}_{\bf 7}^\dag
 \left({\Psi}_{\bf 21}^{(m)}{\Psi}_{\bf \overline{7}}^{(n)}\right)_{\bf 7}
 \nonumber\\
 & \qquad
 -\sum_{m,n=1}^{9}
y_3^{(mn)}
{\Phi}_{\bf 21}
 \left({\Psi}_{\bf \overline{7}}^{(m)}{\Psi}_{\bf \overline{7}}^{(n)}\right)_{\bf \overline{21}}
 -\sum_{m=1}^{3}\sum_{n=1}^{9}
y_4^{(mn)}
{\Phi}_{\bf 7}
 \left({\Psi}_{\bf \overline{7}}^{(m)}{\Psi}_{\bf 48}^{(n)}\right)_{\bf \overline{7}}
 \nonumber\\
 & \qquad
 -\sum_{m,n=1}^{3}
y_5^{(mn)}
{\Phi}_{\bf 21}^{\dagger}
 \left({\Psi}_{\bf 21}^{(m)}{\Psi}_{\bf 48}^{(n)}\right)_{\bf 21}
 - \sum_{m,n=1}^3
y_6^{(mn)}\Phi_{\bf 48}
 \left({\Psi}_{\bf 48}^{(m)}{\Psi}_{\bf 48}^{(n)}\right)_{\bf 48}
 +\mbox{H.c.} \bigg),
\label{Eq:Lagrangian-SU7}
\end{align}
where
$\cancel{D} := \gamma^\mu D_\mu$,
${D}_\mu := \partial_\mu+i{g}{A}_\mu$, and
${F}_{\mu\nu} :=
\partial_\mu{A}_\nu-\partial_\nu{A}_\mu
+i{g}[{A}_\mu,{A}_\nu]$,
with $\mu,\nu=0,1,2,3$;
$y_j^{(mn)}\,(j=1,2,\cdots,6)$ are the coupling constant matrices of the Yukawa
interaction terms;
and $M_M^{(mn)}$ is the Majorana mass matrix.
The scalar potential
$V\left(\left\{{\Phi}_{\bf x}\right\}\right)$
contains quadratic, cubic, and quartic coupling terms,
\begin{align}
V\left(\left\{{\Phi}_{\bf x}\right\}\right)=
V_2\left(\left\{{\Phi}_{\bf x}\right\}\right)
+V_3\left(\left\{{\Phi}_{\bf x}\right\}\right)
+V_4\left(\left\{{\Phi}_{\bf x}\right\}\right),
\label{Eq:Scalar-potential}
\end{align}
where ${\bf x}={\bf 7,21,35,48}$.
(For Lie groups, see, e.g., Refs.~\cite{Yamatsu:2015gut,Feger:2019tvk} and
the scalar potential can also be calculated by using Mathematica packages such as 
GroupMath \cite{Fonseca:2020vke} and
Sym2Int\cite{Fonseca:2017lem,Fonseca:2019yya}.)
The quadratic terms of the scalar potential are given by
\begin{align}
V_2\left(\left\{{\Phi}_{\bf x}\right\}\right)=
\sum_{{\bf y}={\bf 7},{\bf 21},{\bf 35}}
m_{\bf y}^2\left|\Phi_{\bf y}\right|^2
+\frac{1}{2}m_{\bf 48}^2\Phi_{\bf 48}\Phi_{\bf 48},
\label{Eq:Scalar-potential-V2}
\end{align}
where each $m_{\bf y}^2$ stands for the squared mass of the scalar field
$\Phi_{\bf y}$.
The cubic terms of the scalar potential are given by
\begin{align}
V_3\left(\left\{{\Phi}_{\bf x}\right\}\right)=&
{\Big(} \xi_{1}\Phi_{\bf 7}\Phi_{\bf 21}\Phi_{\bf 35}^\dag
+\xi_{2}\Phi_{\bf 21}\Phi_{\bf 21}\Phi_{\bf 35}
+\mbox{H.c.}{\Big)}
\nonumber\\
&
+\sum_{{\bf y}={\bf 7},{\bf 21},{\bf 35}}
\xi_{\bf y}'\Phi_{\bf 48}\left|\Phi_{\bf y}\right|^2
+\xi_{{\bf 48}}'\Phi_{\bf 48}\Phi_{\bf 48}\Phi_{\bf 48},
\label{Eq:Scalar-potential-V3}
\end{align}
where the $\xi$s and $\xi'$s are parameters with the dimension of mass. 
The quartic terms of the scalar potential are given by
\begin{align}
V_4\left(\left\{{\Phi}_{\bf x}\right\}\right)&=
{\Big(} \lambda_{1}\Phi_{\bf 7}\Phi_{\bf 21}\Phi_{\bf 35}^\dag\Phi_{\bf 48}
+\lambda_{2}\Phi_{\bf 21}\Phi_{\bf 21}\Phi_{\bf 35}\Phi_{\bf 48}
+\lambda_{3}\Phi_{\bf 7}\Phi_{\bf 35}\Phi_{\bf 35}\Phi_{\bf 48}
+\lambda_{4}\Phi_{\bf 7}\Phi_{\bf 7}\Phi_{\bf 21}^\dag\Phi_{\bf 48}
\nonumber\\
&\hspace{1em}
+\lambda_{5}\Phi_{\bf 7}\Phi_{\bf 21}\Phi_{\bf 21}\Phi_{\bf 21}
+\lambda_{6}\Phi_{\bf 7}\Phi_{\bf 35}\Phi_{\bf 21}^\dag\Phi_{\bf 21}^\dag
+\lambda_{7}\Phi_{\bf 7}^\dag\Phi_{\bf 21}\Phi_{\bf 35}\Phi_{\bf 35}
+\mbox{H.c.} {\Big)}
\nonumber\\
&\hspace{1em}
+\sum_{{\bf y,z}={\bf 7},{\bf 21},{\bf 35}}
 \lambda_{{\bf y},{\bf z}}
\left|\Phi_{\bf y}\right|^2\left|\Phi_{\bf z}\right|^2
+
\sum_{{\bf y}={\bf 7},{\bf 21},{\bf 35}}
\lambda_{{\bf y}}
\left|\Phi_{\bf y}\right|^2\Phi_{\bf 48}\Phi_{\bf 48}
+\lambda_{{\bf 48}}
\Phi_{\bf 48}\Phi_{\bf 48}\Phi_{\bf 48}\Phi_{\bf 48},
\label{Eq:Scalar-potential-V4}
\end{align}
where the $\lambda$s are dimensionless coupling constants, and
we have omitted duplicated terms. 
There are two invariant terms in each of 
$\Phi_{\bf 7}\Phi_{\bf 21}\Phi_{\bf 35}^\dag\Phi_{\bf 48}$,
$|\Phi_{\bf 35}|^4$, $|\Phi_{\bf 21}|^4$, $(\Phi_{\bf 48})^4$,
$|\Phi_{\bf 7}|^2(\Phi_{\bf 48})^2$,
$|\Phi_{\bf 7}|^2|\Phi_{\bf 21}|^2$, and
$|\Phi_{\bf 7}|^2|\Phi_{\bf 35}|^2$.
There are three invariant terms in each of
$|\Phi_{\bf 21}|^2|\Phi_{\bf 35}|^2$,
$|\Phi_{\bf 35}|^2(\Phi_{\bf 48})^2$, and 
$|\Phi_{\bf 21}|^2(\Phi_{\bf 48})^2$.

\subsection{Symmetry breaking patterns}
\label{Sec:Symmetry-breaking-patterns}

\begin{figure}[tbh]
\begin{center}
\includegraphics[bb=0 0 266 78,height=2.3cm]{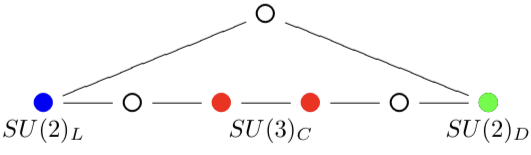}
\includegraphics[bb=0 0 397 122,height=2.3cm]{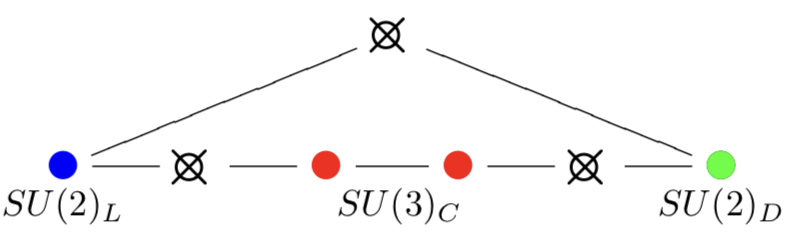}
\end{center}
\caption{\small 
 The (extended) Dynkin diagrams of $SU(7)$ and one of the subgroups 
 $G_{\rm SM}\times SU(2)_D(\times U(1)_X)$ are shown
 in the left and right plots, respectively.
 The red, blue, and green circles stand for the nodes of $SU(3)_C$, $SU(2)_L$, and $SU(2)_D$,
 respectively.
 {\Large$\circ\hspace{-0.65em}\times$} stands for a deleted node
 and the number of remaining $U(1)$ groups is equal to the number of
 {\Large$\circ\hspace{-0.65em}\times$} minus one.
}
\label{Figure:SU7-Dykin-Diagram}
\end{figure}

\begin{figure}[tbh]
\begin{center}
\includegraphics[bb=0 0 389 347,height=7cm]{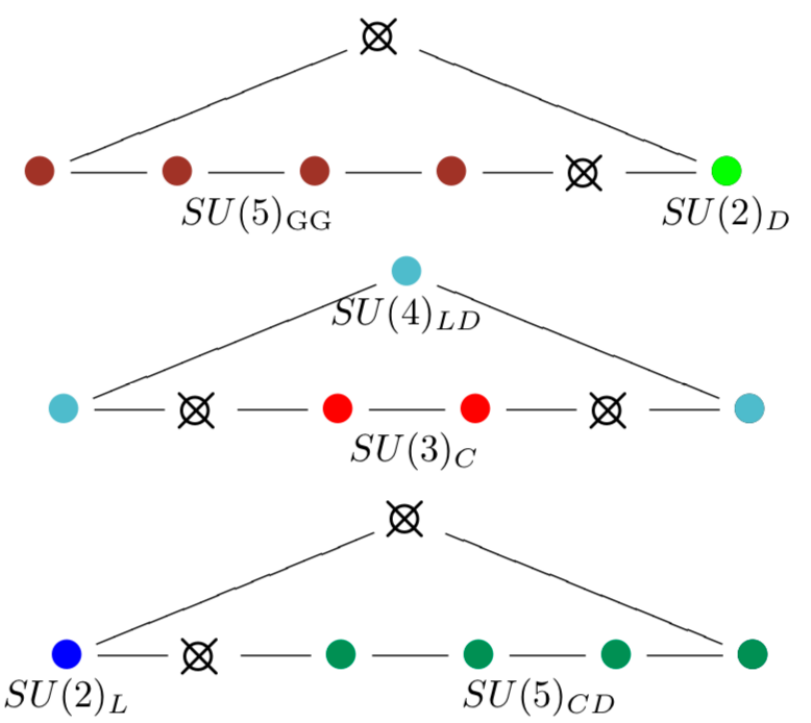}
\end{center}
\caption{\small 
 The (extended) Dynkin diagrams of the maximal subgroups of $SU(7)$
 that include $G_{\rm SM}\times SU(2)_D$, in the same plotting scheme as in Fig.~\ref{Figure:SU7-Dykin-Diagram}. 
 Circles of the same color belong to the same gauge group.
}
\label{Figure:SU7-Maximal-Subgroup}
\end{figure}

The $SU(7)$ GUT gauge group has three different maximal subgroups
preserving the $G_{\rm SM}\times SU(2)_D$ group: 
\begin{align}
 SU(7)&\utxtarrow{}
 \begin{cases}
  G_{521} := SU(5)_{\rm GG}\times SU(2)_D\times U(1)_X \\
  G_{341} := SU(3)_C\times SU(4)_{LD}\times U(1)_a \\
  G_{521}' := SU(5)_{CD}\times SU(2)_L\times U(1)_\alpha\\
 \end{cases}
 \nonumber\\
 &\utxtarrow{}
 \begin{cases}
  \underbrace{SU(3)_C\times SU(2)_L\times U(1)_Y}_{{\rm subgroups\ of}\ SU(5)_{\rm GG}}\times SU(2)_D\times U(1)_X \\
  SU(3)_C\times 
   \underbrace{SU(2)_L\times SU(2)_D\times U(1)_b}_{{\rm subgroups\ of}\ SU(4)_{\rm LD}}
  \times U(1)_a \\
  \underbrace{SU(3)_C\times SU(2)_D\times U(1)_\chi}_{{\rm subgroups\ of}\ SU(5)_{\rm CD}} 
\times SU(2)_L\times U(1)_\alpha\\
 \end{cases}
 \nonumber\\
 &
 =SU(3)_C\times SU(2)_L\times U(1)_Y\times SU(2)_D\times U(1)_X,
\label{Eq:SB-pattern}
\end{align}
where $SU(5)_{\rm GG}$ is the $SU(5)$ grand unified gauge group 
proposed by Georgi and Glashow in Ref.~\cite{Georgi:1974sy}, $SU(5)_{\rm CD}$ encompasses 
the SM color gauge group $SU(3)_C$ and the dark gauge group $SU(2)_D$,
the above $U(1)$ pairs (i.e., $U(1)_a \times U(1)_b$ and $U(1)_\alpha \times U(1)_\chi$) can be rearranged to render $U(1)_Y \times U(1)_X$. 
The above subgroup structure can be easily checked by the (extended)
Dynkin diagram of $SU(7)$ shown in
Figs.~\ref{Figure:SU7-Dykin-Diagram} and
\ref{Figure:SU7-Maximal-Subgroup}.

The symmetry-breaking schemes in Eq.~(\ref{Eq:SB-pattern}) can be
realized by the nonvanishing VEVs of scalar fields in the corresponding
adjoint representation. To check further symmetry-breaking patterns, we
need to know the branching rules of groups and their subgroups.
Many representations of $SU(N)$
$(N=2,3,4,5,7)$ and their branching rules:
$SU(7)\supset SU(5)\times SU(2)\times U(1)$,
$SU(7)\supset SU(4)\times SU(3)\times U(1)$,
$SU(5)\supset SU(3)\times SU(2)\times U(1)$, and 
$SU(4)\supset SU(3)\times U(1)$ can be found 
in Ref.~\cite{Yamatsu:2015gut}.

In the following, we assume the existence of four mass scales:
an electroweak scale denoted by $M_{\rm EW}$, an $SU(2)_D$ breaking scale denoted by $M_D$,
an intermediate scale denoted by $M_I$, and a unification scale denoted by $M_U$.
The mass scales $M_I$ and $M_U$ correspond to the VEVs of 
components of 
$\Phi_{\bf 48}$ and $\Phi_{\bf 21}$ or 
$\Phi_{\bf 21}$ and $\Phi_{\bf 48}$, respectively. 
The VEVs of components of $\Phi_{\bf 48}$ and $\Phi_{\bf 21}$ are
responsible for breaking $SU(7)$ to $G_{\rm SM}\times SU(2)_D$.
The VEV of other components of  $\Phi_{\bf 7}$ and $\Phi_{\bf 48}$ are
responsible for breaking $SU(2)_D$.
Under the above assumptions, we have the following three symmetry
breaking schemes:
\begin{align}
 SU(7)
 &\utxtarrow{\langle{\Phi}_{\bf 21}\rangle\not=0}
 G_{52}=SU(5)_{\rm GG}\times SU(2)_D\ \ \ \mbox{at}\ \mu=M_U
 \nonumber\\
 &\utxtarrow{\langle{\Phi}_{\bf 48}\rangle\not=0}
 G_{\rm SM}\times SU(2)_D\ \ \ \mbox{at}\ \mu=M_I
  \nonumber\\
 &\utxtarrow{\langle{\Phi}_{\bf 7}\rangle\not=0}
\ G_{\rm SM}\ \ \ \mbox{at}\ \mu=M_D
  \nonumber\\
 &
 \utxtarrow{\langle{\Phi}_{\bf 35}\rangle\not=0}
 SU(3)_C\times U(1)_{\rm EM}\ \ \ \mbox{at}\ \mu=M_{\rm EW},
 \label{Eq:SU7-Symmetry-breaking-1} 
\end{align}
and
\begin{align}
 SU(7)
 &\utxtarrow{\langle{\Phi}_{\bf 48}\rangle\not=0}
 \begin{cases}
 G_{341}=SU(3)_C\times SU(4)_{LD}\times U(1)_a \medskip\\
 G_{521}'=SU(5)_{\rm CD}\times SU(2)_L\times U(1)_\alpha\\
 \end{cases}
 \ \  \mbox{at}\ \mu=M_U
 \nonumber\\
 &\utxtarrow{\langle{\Phi}_{\bf 21}\rangle\not=0}
 G_{\rm SM}\times SU(2)_D\ \ \ \mbox{at}\ \mu=M_I
  \nonumber\\
 &\utxtarrow{\langle{\Phi}_{\bf 7}\rangle\not=0}
\ G_{\rm SM} \ \ \mbox{at}\ \mu=M_D
  \nonumber\\
 &
 \utxtarrow{\langle{\Phi}_{\bf 35}\rangle\not=0}
 SU(3)_C\times U(1)_{\rm EM}\ \ \ \mbox{at}\ \mu=M_{\rm EW},
 \label{Eq:SU7-Symmetry-breaking-2} 
\end{align}
where $M_D$ is assumed to be slightly larger than $M_{\rm EW}$, and 
$M_U$ and $M_I$ may be comparable scales and much larger than $M_D$. 
We will see in Sec.~\ref{Sec:RGE} that
for the symmetry-breaking patterns
$SU(7)\to G_{52}\to G_{\rm SM}\times SU(2)_D\to G_{\rm SM}
\to SU(3)_C\times U(1)_{\rm EM}$ 
and $SU(7)\to G_{341}\to G_{\rm SM}
\to SU(3)_C\times U(1)_{\rm EM}$, 
the condition of the gauge coupling unification is not satisfied. 
Therefore, we will mainly focus on the symmetry-breaking pattern
$SU(7)\to G_{521}'\to G_{\rm SM}
\to SU(3)_C\times U(1)_{\rm EM}$.

\begin{table}[htb]
\begin{center}
\begin{tabular}{|c||c|c|c|c|c|c|c|c|c|c|}\hline
 \rowcolor[gray]{0.8}
 &\multicolumn{6}{|c|}{${\Psi}_{\bf 21}^{(m)}$}
 &\multicolumn{3}{|c|}{${\Psi}_{\bf \overline{7}}^{(n)}$}
 \\\hline
 $SU(7)$
 &\multicolumn{6}{|c|}{${\bf 21}$}
 &\multicolumn{3}{|c|}{${\bf \overline{7}}$}
 \\\hline\hline
 \rowcolor[gray]{0.8}
 &\multicolumn{2}{|c|}{$\psi_{{\bf (5,2)}(-3)}$}
 &\multicolumn{3}{|c|}{$\psi_{{\bf (10,1)}(4)}$}
 &\multicolumn{1}{|c|}{$\psi_{{\bf (1,1)}(-10)}$}
 &\multicolumn{2}{|c|}{$\psi_{{\bf (\overline{5},1)}(-2)}$}
 &\multicolumn{1}{|c|}{$\psi_{{\bf (1,2)}(5)}$}
 \\\hline
 $G_{521}'$
 &\multicolumn{2}{|c|}{${{\bf (5,2)}(-3)}$}
 &\multicolumn{3}{|c|}{${{\bf (10,1)}(4)}$}
 &\multicolumn{1}{|c|}{${{\bf (1,1)}(-10)}$}
 &\multicolumn{2}{|c|}{${{\bf (\overline{5},1)}(-2)}$}
 &\multicolumn{1}{|c|}{${{\bf (1,2)}(5)}$}
 \\\hline\hline
 \rowcolor[gray]{0.8}
 & $Q$ & $\tilde{L}^{c}$ & $u^{c}$ & $N$  & $\tilde{d}$ 
 & $e^c$
 & $d^{c}$
 & $\tilde{N}$
 & $L$
 \\\hline\hline
 $SU(3)_C$
 & ${\bf 3}$ & ${\bf 1}$& ${\bf \overline{3}}$  & ${\bf 1}$  & ${\bf 3}$  
 & ${\bf 1}$
 & ${\bf \overline{3}}$ & ${\bf 1}$  & ${\bf 1}$
 \\ \hline
 $SU(2)_{L}$
 & ${\bf 2}$ & ${\bf 2}$ & ${\bf 1}$ & ${\bf 1}$  & ${\bf 1}$ 
 & ${\bf 1}$ & ${\bf 1}$
 & ${\bf 1}$ & ${\bf 2}$
 \\ \hline
 $U(1)_{Y}$
 & $+1/6$ & $+1/2$ &$-2/3$ & $0$  & $-1/3$  & $+1$
 & $+1/3$ & $0$ & $-1/2$
 \\\hline
 $SU(2)_{D}$
 & ${\bf 1}$ & ${\bf 2}$ & ${\bf 1}$ & ${\bf 1}$ & ${\bf 2}$
 & ${\bf 1}$ & ${\bf 1}$ & ${\bf 2}$ & ${\bf 1}$
 \\ \hline
 $U(1)_{X}$
 & $+4$ & $-3$ & $+4$  & $-10$ & $-3$  & $+4$
 & $-2$ & $+5$ & $-2$
 \\\hline\hline
 $U(1)_{\alpha}$
 & $-3$ & $-3$ & $+4$  & $+4$ & $+4$  & $-10$
 & $-2$ & $-2$ & $+5$
 \\\hline
 $U(1)_{\chi}$
 & $+2$ & $-3$ & $+4$  & $-6$ & $-1$  & $0$
 & $-2$ & $+3$ & $0$
 \\\hline
\end{tabular}\\[0.5em]
\begin{tabular}{|c||c|c|c|c|c|c|c|c|c|c|c|}\hline
 \rowcolor[gray]{0.8}
 &\multicolumn{11}{|c|}{${\Psi}_{\bf 48}^{(m)}$}
 \\\hline
 $SU(7)$
 &\multicolumn{11}{|c|}{${\bf 48}$}
 \\\hline\hline
 \rowcolor[gray]{0.8}
 &\multicolumn{5}{|c|}{$\psi_{{\bf (24,1)}(0)}$}
 &\multicolumn{1}{|c|}{$\psi_{{\bf (1,3)}(0)}$}
 &\multicolumn{1}{|c|}{$\psi_{{\bf (1,1)}(0)}$}
 &\multicolumn{2}{|c|}{$\psi_{{\bf (5,2)}(7)}$}
 &\multicolumn{2}{|c|}{$\psi_{{\bf (\overline{5},2)}(-7)}$}
 \\\hline
 $G_{521}'$
 &\multicolumn{5}{|c|}{${{\bf (24,1)}(0)}$}
 &\multicolumn{1}{|c|}{${{\bf (1,3)}(0)}$}
 &\multicolumn{1}{|c|}{${{\bf (1,1)}(0)}$}
 &\multicolumn{2}{|c|}{${{\bf (5,2)}(7)}$}
 &\multicolumn{2}{|c|}{${{\bf (\overline{5},2)}(-7)}$}
 \\\hline\hline
 \rowcolor[gray]{0.8}
 & $\tilde{g}$ & $\tilde{W}'$ & $\tilde{B}$
 & $d_D'$ & $d_D^{\prime c}$ 
 & $\tilde{W}$ 
 & $\tilde{B}'$
 & $\tilde{X}$ & $L_D'$
 & $\tilde{X}^{c}$ & $L_D^{\prime c}$
 \\\hline\hline
 $SU(3)_C$
 & ${\bf 8}$ & ${\bf 1}$ & ${\bf 1}$
 & ${\bf 3}$ & ${\bf \overline{3}}$
 & ${\bf 1}$
 & ${\bf 1}$ 
 & ${\bf 3}$ & ${\bf 1}$ 
 & ${\bf \overline{3}}$ & ${\bf 1}$ 
 \\ \hline
 $SU(2)_{L}$
 & ${\bf 1}$ & ${\bf 1}$ & ${\bf 1}$
 & ${\bf 1}$ & ${\bf 1}$
 & ${\bf 3}$ 
 & ${\bf 1}$ 
 & ${\bf 2}$ & ${\bf 2}$ 
 & ${\bf 2}$ & ${\bf 2}$ 
 \\ \hline
 $U(1)_{Y}$
 & $0$ & $0$ & $0$ 
 & $-1/3$  & $+1/3$ 
 & $0$ 
 & $0$ 
 & $-5/6$ & $+1/2$ 
 & $+5/6$ & $-1/2$  
 \\\hline
 $SU(2)_{D}$
 & ${\bf 1}$ & ${\bf 3}$ & ${\bf 1}$ 
 & ${\bf 2}$ & ${\bf 2}$
 & ${\bf 1}$ 
 & ${\bf 1}$
 & ${\bf 1}$ & ${\bf 2}$
 & ${\bf 1}$ & ${\bf 2}$ 
 \\ \hline
 $U(1)_{X}$
 & $0$ & $0$ & $0$
 & $+7$ & $-7$ 
 & $0$ & $0$
 & $0$ & $-7$ 
 & $0$ & $+7$ 
 \\\hline\hline
 $U(1)_{\alpha}$
 & $0$ & $0$ & $0$
 & $0$ & $0$
 & $0$
 & $0$
 & $+7$ & $+7$ 
 & $-7$ & $-7$ 
 \\\hline
 $U(1)_{\chi}$
 & $0$ & $0$ & $0$
 & $+5$ & $-5$
 & $0$
 & $0$
 & $+2$ & $-3$ 
 & $-2$ & $+3$ 
 \\\hline
\end{tabular}
 \caption{\small The content of fermions in the $SU(7)$ model is shown
 in the $G_{521'} := SU(5)_{CD}\times SU(2)_L\times U(1)_\alpha$ basis,
 all belonging to $(1/2,0)$ under $SL(2,\mathbb{C})$.
 }
\label{Tab:Matter_Content-GUT-fermion-G521'}
\end{center}  
\end{table}

\begin{table}[htb]
\begin{center}
\begin{tabular}{|c||c|c|c|c|c|c|c|}\hline
 \rowcolor[gray]{0.8}
 &\multicolumn{1}{|c|}{${\Phi}_{\bf 7}$}
 &${\Phi}_{\bf 21}$
 &${\Phi}_{\bf 35}$
 &\multicolumn{2}{|c|}{${\Phi}_{48}$}
 \\\hline
 $SU(7)$
 &\multicolumn{1}{|c|}{${\bf 7}$}
 &${\bf 21}$
 &${\bf 35}$
 &\multicolumn{2}{|c|}{${\bf 48}$}
 \\\hline\hline
 \rowcolor[gray]{0.8}
 &\multicolumn{1}{|c|}{$\phi_{{\bf (5,1)}(+2)}$}
 &$\phi_{{\bf (10,1)}(+4)}$
 &$\phi_{{\bf (10,2)}(-1)}$
 &$\phi_{{\bf (1,1)}(0)}$
 &$\phi_{{\bf (24,1)}(0)}$
 \\\hline
 $G_{521}'$
 &\multicolumn{1}{|c|}{${{\bf (5,1)}(+2)}$}
 &${{\bf (10,1)}(+4)}$
 &${{\bf (10,2)}(-1)}$
 &${{\bf (1,1)}(0)}$
 &${{\bf (24,1)}(0)}$
 \\\hline\hline
 \rowcolor[gray]{0.8}
 & $\Phi$ & $X$ & $H$ & $\Omega$  &$\Delta$ 
 \\\hline\hline
 $SU(3)_C$
 & ${\bf 1}$ & ${\bf 1}$ & ${\bf 1}$ & ${\bf 1}$& ${\bf 1}$
 \\ \hline
 $SU(2)_{L}$
 & ${\bf 1}$ & ${\bf 1}$ & ${\bf 2}$ & ${\bf 1}$& ${\bf 1}$
 \\ \hline
 $U(1)_{Y}$
 & $0$ & $0$ &$+1/2$ & $0$ & $0$ 
 \\\hline
 $SU(2)_{D}$
 & ${\bf 2}$ & ${\bf 1}$& ${\bf 1}$ & ${\bf 1}$& ${\bf 3}$ 
 \\ \hline
 $U(1)_{X}$
 & $-5$ & $-10$ & $-8$ & $0$& $0$ 
 \\\hline\hline
 $U(1)_1$
 & $+2$ & $+4$ & $-1$ & $0$& $0$ 
 \\\hline
 $U(1)_2$
 & $-3$ & $-6$ & $-6$ & $0$& $0$ 
 \\\hline
\end{tabular}
 \caption{\small The content of scalar fields in the $SU(7)$ model is
 shown in the $G_{521}'$ basis.
The  ${\Phi}_{\bf 7}$,
 ${\Phi}_{\bf 21}$, and 
 ${\Phi}_{\bf 35}$ fields
 are complex scalars, while
 the  ${\Phi}_{\bf 48}$ field
 is a real scalar.
 }
\label{Tab:Matter_Content-GUT-scalar-G521'}
\end{center}  
\end{table}

\begin{table}[htb]
\begin{center}
\begin{tabular}{|c||c|c|c|c|c|}\hline
 \rowcolor[gray]{0.8}
 &\multicolumn{5}{|c|}{${A}_\mu$}
 \\\hline
 $SU(7)$
 &\multicolumn{5}{|c|}{${\bf 48}$}
 \\\hline\hline
 \rowcolor[gray]{0.8}
 &\multicolumn{3}{|c|}{$G_\mu'$}&$W_\mu$&$B_\mu'$
 \\\hline
 $G_{521}'$
 &\multicolumn{3}{|c|}{${\bf (24,1)}(0)$}
 &${\bf (1,3)}(0)$
 &${\bf (1,1)}(0)$
 \\\hline\hline
 \rowcolor[gray]{0.8}
 & $G_\mu$ &$W_\mu'$&$B_\mu$& $W_\mu$ & $B_\mu'$
 \\\hline\hline
 $SU(3)_C$
 & ${\bf 8}$ & ${\bf 1}$ & ${\bf 1}$ & ${\bf 1}$ & ${\bf 1}$
 \\ \hline
 $SU(2)_{L}$
 & ${\bf 1}$ & ${\bf 1}$ & ${\bf 1}$ & ${\bf 3}$ & ${\bf 1}$
 \\ \hline
 $U(1)_{Y}$
 & $0$ & $0$ & $0$ & $0$ & $0$
 \\\hline
 $SU(2)_{D}$
 & ${\bf 1}$ & ${\bf 3}$ & ${\bf 1}$ & ${\bf 1}$ & ${\bf 1}$
 \\ \hline
 $U(1)_{X}$
 & $0$ & $0$ & $0$ & $0$ & $0$
 \\\hline
\end{tabular}
 \caption{\small The content of gauge fields in the $SU(7)$ model is
 shown in the $G_{521}'$ basis.
 }
\label{Tab:Matter_Content-GUT-gauge-G521'}
\end{center}  
\end{table}

Here we consider the breakdown of $SU(7)$ to
$G_{521}'$
by the nonvanishing VEV of the $G_{521}'$ $({\bf 1,1})(0)$ component 
of the $\Phi_{\bf 48}$ field, denoted by $\Omega$,
at the grand unification scale $M_U$. 
\begin{align}
 SU(7)
 &\utxtarrow{v_\Omega\not=0}
 G_{521}',
 \label{Eq:SU7-Symmetry-breaking-MU-G521'}
\end{align}
where $v_\Omega$ stands for the nonvanishing VEV of $\Omega$.
The field content below $\mu=M_U$ is shown in
Tables~\ref{Tab:Matter_Content-GUT-fermion-G521'},
\ref{Tab:Matter_Content-GUT-scalar-G521'}, and
\ref{Tab:Matter_Content-GUT-gauge-G521'}, where 
the $U(1)_{X}$ charge $Q_{X}$ is given by
$U(1)(\subset SU(7)/SU(5)\times SU(2)$) \cite{Yamatsu:2015gut}.
In the tables, the charges $Q_{Y}=-(3Q_\alpha+2Q_\chi)/30$ and
$Q_{X}=(-2Q_\alpha+7Q_\chi)/5$, and
all unlisted components of $G'_{521}$ have
masses of ${\cal O}(M_{U})$ and also all unlisted components of
$G_{\rm SM}\times SU(2)_D\times U(1)_X$ have masses of
$\mathcal{O}(M_{I})$ or $\mathcal{O}(M_{U})$.
At this stage, all fermions are chiral under $G_{521}'$ and
are thus massless. The gauge fields corresponding to the unbroken gauge
symmetry $G_{521}'$ are massless, and the others have masses of ${\cal O}(M_U)$.
The typical masses of scalar fields are expected to be of ${\cal
O}(M_U)$.  Nonetheless, we assume that some scalar fields responsible
for subsequent symmetry breakdowns have accidentally small masses
compared to $M_U$
to realize gauge coupling unification.
Next, $G_{521}'$ is broken to $G_{\rm SM}\times SU(2)_D$ at the
intermediate scale $M_I$ by the VEV of the 
$G_{521}'$ $({\bf 10,1})(+4)$ component 
of the scalar field
$\Phi_{\bf 21}$, denoted by $X$,
\begin{align}
 G_{521}'
 \utxtarrow{v_X\not=0}
 G_{\rm SM}\times SU(2)_D,
 \label{Eq:SU7-Symmetry-breaking-MI-G521'}
\end{align}
where $v_X$ stands for the nonvanishing VEV of $X$.
More specifically,
the VEV of $\Phi_{\bf 21}$ $v_X$ breaks $SU(5)_{CD}\times U(1)_\alpha$
to $G_{\rm SM}\times SU(2)_D$.
Further, $SU(2)_D$ is broken at the scale $M_D$ by
the VEV of the 
$G_{521}'$ $({\bf 5,1})(+2)$ and  $({\bf 24,1})(0)$
components of the scalar fields in $\Phi_{\bf 7}$ and 
$\Phi_{\bf 48}$, denoted respectively by $\Phi$ and $\Delta$,
\begin{align}
 G_{\rm SM}\times SU(2)_D
 \utxtarrow{v_\Phi,v_\Delta\not=0}
 G_{\rm SM},
 \label{Eq:SU7-Symmetry-breaking-MD-G521'}
\end{align}
where $v_\Phi$ and $v_\Delta$ stand for the nonvanishing VEVs of $\Phi$
and $\Omega$, respectively.
Finally, $G_{\rm SM}$ is broken to $SU(3)_C\times U(1)_{\rm EM}$ at
the electroweak scale $M_{\rm EW}$ by the nonzero VEV of the SM Higgs field $H$, which is a linear combination of components of $\Phi_{\bf 7}$ and $\Phi_{\bf 35}$.  In summary, we have
\begin{align}
 SU(7)
 &\utxtarrow{v_\Omega\not=0}
 G_{521}'
 \utxtarrow{v_X\not=0}
 G_{\rm SM}\times SU(2)_D
 \utxtarrow{v_\Phi,v_\Delta\not=0}
 G_{\rm SM}
 \utxtarrow{v\not=0}
 SU(3)_C\times U(1)_{\rm EM},
 \label{Eq:SU7-Symmetry-breaking-MEW-G521'}
\end{align}
where $v$ stands for the nonvanishing VEV of $H$.
The up-type quarks, down-type quarks, and charged leptons become
massive via the first and second Yukawa coupling terms in
Eq.~(\ref{Eq:Lagrangian-SU7}) and the VEV of the Higgs field $H$ in Table \ref{Tab:Matter_Content-GUT-scalar-G521'}. 

With the particle content of the model defined above, the unification scale $M_U$ and the intermediate scale $M_I$
will be determined by gauge coupling unification condition using the 
renormalization group equations (RGEs) for the gauge coupling constants
in the next section.

\subsection{Lagrangian without heavy particles}
\label{Sec:Lagrangian-low-energy}

We have so far described the Lagrangian and symmetry-breaking patterns
of the $SU(7)$ GUT model, and we now consider a non-Abelian pNGB DM
model as a low-energy effective theory of the GUT model.
The field content of this low-energy effective theory includes the field
content of the non-Abelian pNGB DM model discussed in
Ref.~\cite{Otsuka:2022zdy} and additional vector-like fermions required
by the grand unification. 

The low-energy effective theory consists of the SM gauge fields, 
an $SU(2)_D$ gauge field ${W}_\mu^{\prime a}$ $(a=1,2,3)$,
a complex scalar field in ${\bf 2}$ of $SU(2)_D$ denoted by
$\Phi$ contained in $\Phi_{\bf 7}$,
a real scalar field in ${\bf 3}$ of $SU(2)_D$ denoted by
$\Delta$ contained in $\Phi_{\bf 48}$,
and vectorlike fermions.
The field content in the non-Abelian pNGB DM model is summarized in
Table~\ref{Tab:Matter_content}.
In the low-energy regime, this model has the same field
content as in Ref.~\cite{Otsuka:2022zdy} along with the additional vectorlike
fermions.
Six $\tilde{N}^{(n)}$ and three $\tilde{D}^{c(m)}$, six $L^{(n)}$ and
three $\tilde{L}^{c(m)}$, and three $N^{(m)}$ become
massive through the VEVs $v_\Phi$ and
$v_\Delta$. 
We assume that the SM Higgs boson $H$ as $({\bf 1,2})(+1/2)$ under $G_{\rm SM}$ is a linear combination of components in
$\Phi_{\bf 35}$ and $\Phi_{\bf 7}$.

\begin{table}[thb]
\begin{center}
\begin{tabular}{|c||c|c|c|c|c||c|c|c|c|c|c|c|}\hline
 \rowcolor[gray]{0.8}
 & $Q^{(m)}$ & $u^{c(m)}$ & $d^{c(n)}$
 & $L^{(n)}$ & $e^{c(m)}$
 & $\tilde{d}^{(m)}$ 
 & $\tilde{L}^{c(m)}$
 & $N^{(m)}$
 & $\tilde{N}^{(n)}$
 \\\hline\hline
 $SU(3)_{C}$
 & ${\bf 3}$ & ${\bf \overline{3}}$ & ${\bf \overline{3}}$
 & ${\bf 1}$ & ${\bf 1}$ 
 & ${\bf 3}$ & ${\bf 1}$ & ${\bf 1}$ & ${\bf 1}$ 
 \\ \hline
 $SU(2)_{W}$
 & ${\bf 2}$ & ${\bf 1}$ & ${\bf 1}$
 & ${\bf 2}$ & ${\bf 1}$ 
 & ${\bf 1}$ & ${\bf 2}$ & ${\bf 1}$ & ${\bf 1}$
 \\ \hline
 $U(1)_{Y}$
 & $+1/6$ & $-2/3$ & $+1/3$
 & $-1/2$ & $+1$  
 & $-1/3$  & $+1/2$  & $0$ & $0$
 \\\hline
 $SU(2)_{D}$
 & ${\bf 1}$ & ${\bf 1}$ & ${\bf 1}$
 & ${\bf 1}$ & ${\bf 1}$
 & ${\bf 2}$ & ${\bf 2}$ & ${\bf 1}$ & ${\bf 2}$ 
  \\ \hline
\end{tabular}\\[0.5em]
\begin{tabular}{|c||c|c|c||c|c|c|c|}\hline
 \rowcolor[gray]{0.8}
 & $H$ 
 & $\Phi$
 & $\Delta$
 & $G_{\mu\nu}$
 & $W_{\mu\nu}$
 & $B_{\mu\nu}$
 & $W_{\mu\nu}'$
 \\\hline\hline
 $SU(3)_{C}$
 & ${\bf 1}$ & ${\bf 1}$ & ${\bf 1}$ 
 & ${\bf 8}$ & ${\bf 1}$ & ${\bf 1}$ & ${\bf 1}$ 
 \\ \hline
 $SU(2)_{W}$
 & ${\bf 2}$ & ${\bf 1}$ & ${\bf 1}$ 
 & ${\bf 1}$ & ${\bf 3}$ & ${\bf 1}$ & ${\bf 1}$ 
 \\ \hline
 $U(1)_{Y}$
 & $+1/2$  & $0$  & $0$
 & $0$ & $0$ & $0$ & $0$
 \\\hline
 $SU(2)_{D}$
 & ${\bf 1}$ & ${\bf 2}$ & ${\bf 3}$
 & ${\bf 1}$ & ${\bf 1}$ & ${\bf 1}$ & ${\bf 3}$ 
 \\ \hline
\end{tabular}
 \caption{\small The field content in the pNGB DM model 
 is shown in the $G_{\rm SM}\times SU(2)_D$ basis,
 where the fermions belong to $(1/2,0)$ under $SL(2,\mathbb{C})$.
 $m=1,2,3$; $n=1,2,...,9$.
 }
\label{Tab:Matter_content}
\end{center}  
\end{table}

The Lagrangian of the low-energy effective theory is given by
\begin{align}
 {\cal L}&=
  -\frac{1}{2}\mbox{tr}
 \left[{G}_{\mu\nu}{G}^{\mu\nu}\right]
  -\frac{1}{2}\mbox{tr}
 \left[{W}_{\mu\nu}{W}^{\mu\nu}\right]
  -\frac{1}{4}
 {B}_{\mu\nu}{B}^{\mu\nu}
  -\frac{1}{2}\mbox{tr}
 \left[{W}_{\mu\nu}^{\prime}{W}^{\prime\mu\nu}\right]
\allowdisplaybreaks[1]\nonumber\\
 &\hspace{1em}
 +\left({D}_\mu H\right)^\dag
 \left({D}^\mu H\right)
 +\left({D}_\mu{\Phi}\right)^\dag
 \left({D}^\mu{\Phi}\right)
 +\frac{1}{2}
 \mbox{tr}\left[
 \left({D}_\mu\Delta\right)
 \left({D}^\mu\Delta\right)\right]
 -{\cal V}\left(H,\Phi,\Delta\right)
\allowdisplaybreaks[1]\nonumber\\
 &\hspace{1em}
 +\sum_{m=1}^3
 \bigg(
 \overline{Q^{(m)}}i\cancel{{D}}Q^{(m)}
 +\overline{u^{c(m)}}i\cancel{{D}}u^{c(m)}
 +\overline{e^{c(m)}}i\cancel{{D}}e^{c(m)}
\nonumber\\
 &\hspace{5em}
 +\overline{\tilde{d}^{(m)}}i\cancel{{D}}\tilde{d}^{(m)}
 +\overline{\tilde{L}^{c(m)}}i\cancel{{D}}\tilde{L}^{c(m)}
 +\overline{N^{(m)}}i\cancel{{D}}N^{(m)}
 \bigg)
\allowdisplaybreaks[1]\nonumber\\
 &\hspace{1em}
 +\sum_{n=1}^9
 \left(
 \overline{d^{c(n)}}i\cancel{{D}}d^{c(n)}
 +\overline{L^{(n)}}i\cancel{{D}}L^{(n)}
 +\overline{\tilde{N}^{(n)}}i\cancel{{D}}\tilde{N}^{(n)}
 \right)
\allowdisplaybreaks[1]\nonumber\\
 &\hspace{1em}
 +\sum_{m,n=1}^9M_{\tilde{N}}^{(mn)}\tilde{N}^{(m)}\tilde{N}^{(n)}
 +\sum_{m,n=1}^9M_N^{(mn)}N^{(m)}N^{(n)}
\allowdisplaybreaks[1]\nonumber\\
 &\hspace{1em}
 -\bigg(
 \sum_{m,n=1}^3
 y_u^{(mn)}Q^{(m)}u^{c(n)} H
 +\sum_{m=1}^3\sum_{n=1}^9
  y_d^{(mn)}Q^{(m)}d^{c(n)} H^\dag
 \nonumber\\
 &\hspace{5em}
 +\sum_{m=1}^9\sum_{n=1}^3
 y_e^{(mn)}L^{(m)}e^{c(n)}H^\dag
 +\sum_{m=1}^3\sum_{n=1}^9
  y_\nu^{(mn)}N^{(m)}L^{(n)} H
 +\mbox{H.c.}
 \bigg)
\allowdisplaybreaks[1]\nonumber\\
 &\hspace{1em}
 -\sum_{m=1}^3\sum_{n=1}^9
 \left(
  y_{Dd}^{(mn)}\tilde{d}^{(m)}d^{c(n)} i\sigma_2\Phi^*
 +y_{DL}^{(mn)}\tilde{L}^{c(m)}L^{(n)} i\sigma_2\Phi^*
 +y_{DN}^{(mn)}N^{(m)}\tilde{N}^{(n)} i\sigma_2\Phi^*
 +\mbox{H.c.}\right)
\allowdisplaybreaks[1]\nonumber\\
 &\hspace{1em}
 -\sum_{m=1}^3\sum_{n=1}^9
 \left(
  y_{Dd}^{\prime(mn)}\tilde{d}^{(m)}d^{c(n)} \Phi
 +y_{DL}^{\prime(mn)}\tilde{L}^{c(m)}L^{(n)} \Phi
 +y_{DN}^{\prime(mn)}N^{(m)}\tilde{N}^{(n)} \Phi
 +\mbox{H.c.}\right)
\allowdisplaybreaks[1]\nonumber\\
 &\hspace{1em}
 -\sum_{m,n=1}^9
 \left(
  y_{D\Delta}^{(mn)}\tilde{N}^{(m)}\tilde{N}^{(n)} \Delta
 +\mbox{H.c.}\right),
\label{Eq:Lagrangian}
\end{align}
where ${F}_{\mu\nu}=
\partial_\mu{F}_\nu-\partial_\nu{F}_\mu+i{g}[{F}_\mu,{F}_\nu]$ with $F=G,W,B,W'$ and $g$ is the corresponding gauge coupling constant.
The primed Yukawa coupling matrices $y_{Dd}^{\prime(mn)}$, $y_{DL}^{\prime(mn)}$ and $y_{DN}^{\prime(mn)}$ do not appear at a renormalizable level, but are generated through higher-order operators through $y_4^{(mn)}$ and $y_5^{(mn)}$, e.g., 
\begin{align}
y_{Dd}^{\prime(mn)} \simeq  y_4^{(m k)}
(M_{\Psi_{\bf 48}}^{-1})^{(k\ell)}y_5^{(\ell n)} v_X^{}.
\end{align}
We will see later in Sec.~4 that the simultaneous conditions \{$y_4^{(mn)}=0$ or $y_5^{(mn)}=0$\} and $\kappa_1=\kappa_2=0$ guarantee the stability of DM. 
We assume that the field content for the scalar fields in the model below
$M_I$ is the same as that in Ref.~\cite{Otsuka:2022zdy}, so the scalar
potential in the model must be the same as that in
Ref.~\cite{Otsuka:2022zdy}
when we write down a potential that contains only renormalizable terms.
The scalar potential 
${\cal V}\left(H,\Phi,\Delta\right)$ that 
contains quadratic, cubic, and quartic coupling terms is given by
\begin{align}
 {\cal V}(H,\Phi,\Delta)&=
 -\mu_H^2H^\dag H
 -\mu_{\Phi}^2\Phi^\dag\Phi
 -\frac{1}{2}\mu_{\Delta}^2\mbox{Tr}\left(\Delta^2\right)
 \nonumber\\
 &\hspace{1em}
 +
 \sqrt{2}
 \left(
  (\kappa_1+i\kappa_2)
 (i\sigma_2\Phi^*)^\dag\Delta\Phi
 + (\kappa_1-i\kappa_2)
 {\Phi}^\dag\Delta(i\sigma_2\Phi^*)
 \right)
 +2\sqrt{2}\kappa_3\Phi^\dag\Delta\Phi
 \nonumber\\
 &\hspace{1em}
 +\lambda_{H}\left(H^\dag H\right)^2
 +\lambda_{\Phi}\left(\Phi^\dag \Phi\right)^2
 +\frac{1}{4}\lambda_{\Delta}\mbox{Tr}\left(\Delta^2\right)^2
 \nonumber\\
 &\hspace{1em}
 +\lambda_{H\Phi} \left(H^\dag H\right) \left(\Phi^\dag \Phi\right)
 +\lambda_{H\Delta}\left(H^\dag H\right)
 \mbox{Tr}\left(\Delta^2\right)
 +\lambda_{\Phi\Delta}
 \left(\Phi^\dag \Phi\right)
  \mbox{Tr}\left(\Delta^2\right),
\label{Eq:Potential-scalar}
\end{align}
where 
$\mu_H^{2}$, $\mu_\Phi^{2}$, and $\mu_\Delta^{2}$
are real parameters with mass dimension 2, 
$\kappa_a~(a=1,2,3)$ are real parameters with mass dimension 1,
and $\lambda_{H}$, $\lambda_{\Phi}$, $\lambda_{\Delta}$
$\lambda_{H\Phi}$, $\lambda_{H\Delta}$, and $\lambda_{\Phi\Delta}$
are dimensionless real parameters.
We use the following component parameterization for the dark scalar fields:
\begin{align}
 \Delta&=
 \frac{1}{\sqrt{2}}\left(
 \begin{array}{cc}
  \eta_3&\eta_1-i\eta_2 \\
  \eta_1+i\eta_2 &-\eta_3\\
 \end{array}
 \right),\ \ \
 \Phi=
 \frac{1}{\sqrt{2}}\left(
 \begin{array}{c}
  \phi_1+i\phi_2\\
  \phi_3+i\phi_4\\
 \end{array}
 \right).
 \label{Eq:scalar-component}
\end{align}
Under the $SU(2)_D$ transformations, $\Phi(x)$ and $\Delta(x)$
behave as 
\begin{align}
 &\Phi(x)\to U(x) \Phi(x),\ \
 \Delta(x)\to U(x)\Delta(x) U(x)^{\dag},\
\end{align}
where $U(x)=\mbox{exp}\left[i\theta_a(x)\frac{\sigma_a}{2}\right]$ denotes an $SU(2)_D$ unitary transformation,
$\theta_a(x)~(a=1,2,3)$ are the parameters of the $SU(2)_D$ gauge
transformation, and $\sigma_a$ stand for the Pauli matrices.

Next, we consider the parameter relation between $SU(7)$ and 
$G_{\rm SM}\times SU(2)_D$ in the scalar sector.
We need to write down terms that include only the degrees of freedom
in the low-energy regime and the nonvanishing VEVs.  We have
\begin{align}
\Phi_{\bf 7} &\to h_7=h\sin{\theta_h},\ \phi_7=\phi\sin\theta_{\phi},
\nonumber\\
\Phi_{\bf 21} &\to v_X,
\nonumber\\
\Phi_{\bf 35} &\to h_{35}=h\cos{\theta_h},\ \phi_{35}=\phi\cos\theta_{\phi},
\nonumber\\
\Phi_{\bf 48} &\to v_\Omega,\ \Delta,
\label{Eq:Low-energy-DoF}
\end{align}
where $\phi_7$ and $h_7$ belong to 
$({\bf 1,1})(0)({\bf 2})(-5)$ and $({\bf 1,2})(+1/2)({\bf 2})(+2)$ 
of
$G_{32121} :=
SU(3)_C\times SU(2)_L\times U(1)_Y\times SU(2)_D\times U(1)_X$,
$\phi_{35}$ and $h_{35}$ belong to 
$({\bf 1,1})(0)({\bf 2})(-1)$ and $({\bf 1,2})(+1/2)({\bf 2})(-8)$,
and $\Delta$ belongs to 
$({\bf 1,1})(0)({\bf 3})(0)$.
Substituting the relations in Eq.~(\ref{Eq:Low-energy-DoF}) into
Eq.~(\ref{Eq:Scalar-potential}),
the scalar potential can be written as
\begin{align}
V\left(H,\Phi,\Delta;v_X,v_\Omega\right)=
V_2\left(H,\Phi,\Delta;v_X,v_\Omega\right)
+V_3\left(H,\Phi,\Delta;v_X,v_\Omega\right)
+V_4\left(H,\Phi,\Delta;v_X,v_\Omega\right).
\label{Eq:Scalar-potential-LE}
\end{align}
By comparing terms in
Eqs.~(\ref{Eq:Scalar-potential-LE}) and (\ref{Eq:Potential-scalar}),
we can write the low-energy parameters in
Eq.~(\ref{Eq:Potential-scalar}) in terms of the $SU(7)$ GUT parameters in Eq.~(\ref{Eq:Scalar-potential})
at tree level.

We will check below how the parameters in the Lagrangian of $SU(7)$
relate to the parameters in the Lagrangian of $G_{\rm SM}\times SU(2)_D$.
First, we consider the Yukawa interaction terms between fermions and 
the Higgs bosons.
Suppose the SM Higgs boson is realized as a linear combination of
the scalar fields in $({\bf 1,2})(+1/2)$ of $G_{\rm SM}$
\begin{align}
\left(
\begin{array}{c}
h_{35}\\
h_{7}\\
\end{array}
\right) 
=
\left(
\begin{array}{cc}
\cos{\theta_h}&-\sin{\theta_h}\\
\sin{\theta_h}&\cos{\theta_h}\\
\end{array}
\right)
\left(
\begin{array}{c}
h\\
h'\\
\end{array}
\right),
\end{align}
where $h_{35}$ and $h_{7}$ are the component fields of  
the scalars $\Phi_{\bf 35}$ and $\Phi_{\bf 7}$, and 
${\theta_h}$ is the mixing angle of the Higgs bosons. We assume that 
the mass of $h$ is $m_h=125$\,GeV and the mass of $h'$ is of ${\cal O}(M_U)$. 
The Yukawa coupling constants of the SM fermions are given by 
using the coupling constants of $SU(7)$ as follows:
\begin{align}
y_{u}^{(mn)}=y_1^{(mn)}\cos{\theta_h},\
y_{d}^{(mn)}=y_2^{(mn)}\sin{\theta_h},\
y_{e}^{(mn)}=(y_2^{(mn)})^T\sin{\theta_h},\
y_{\nu}^{(mn)}=y_2^{(mn)}\sin{\theta_h}.
\end{align}
Suppose the scalar boson in ${\bf 2}$ of $SU(2)_D$ is realized as a
linear combination of the scalar fields 
\begin{align}
\left(
\begin{array}{c}
\phi_{35}\\
\phi_{7}\\
\end{array}
\right) 
=
\left(
\begin{array}{cc}
\cos\theta_{\phi}&-\sin\theta_{\phi}\\
\sin\theta_{\phi}&\cos\theta_{\phi}\\
\end{array}
\right)
\left(
\begin{array}{c}
\phi\\
\phi'\\
\end{array}
\right),
\end{align}
where $\phi_{35}$ and $\phi_{7}$ are the component fields of  
the scalars $\Phi_{\bf 35}$ and $\Phi_{\bf 7}$, and 
$\theta_{\phi}$ is the mixing angle.
The other Yukawa coupling constants are given by
\begin{align}
y_{Dd}^{(mn)}=y_2^{(mn)}\sin\theta_{\phi},\
y_{DL}^{(mn)}=y_2^{(mn)}\sin\theta_{\phi},\
y_{DN}^{(mn)}=y_2^{(mn)}\sin\theta_{\phi}.
\end{align}

We now consider the masses of the $\tilde{N}^{(m)}$ and $N^{(m)}$
fermions, which arise from several mass terms.
First, the $\tilde{N}^{(m)}$ fermions have a ``Dirac mass'' term
generated by the VEV of $X$:
\begin{align}
 M_{\tilde{N}}^{(mn)}&=y_3^{(mn)}v_X,
\label{Eq:M_D-v_X}
\end{align}
where $v_X$ is the VEV of the scalar field $X$ and of ${\cal O}(M_I)$.
At least one eigenvalue of $M_{\tilde{N}}^{(mn)}$ in
Eq.~(\ref{Eq:M_D-v_X}) is 
zero because the number of $\tilde{N}^{(m)}$ is odd and
$M_{\tilde{N}}^{(mn)}$ is 
antisymmetric in the indices $m$ and $n$. 
Second, the $\tilde{N}^{(m)}$ fermions have a Majorana mass term
generated by the VEV of $\Delta$. 
The Yukawa coupling constant $y_{D\Delta}^{(mn)}$ of
$\tilde{N}^{(m)}\tilde{N}^{(n)}\Delta$ can be written as the following
effective operator:
\begin{align}
\frac{v_X}{\Lambda}\tilde{N}^{(m)}\tilde{N}^{(n)}\Delta,
\end{align}
When the real component of $\Phi$ develops the VEV, it induces at least
the following effective coupling constant 
\begin{align}
y_{D\Delta}^{(mn)}&\sim
y_3^{(mn)}\left(m_{X}^2\right)^{-1}\xi_{\bf 21}'v_X
+y_3^{(mn)}\left(m_{X}^2\right)^{-1}\lambda_{\bf 21}v_Xv_\Omega
\nonumber\\
&\ \ 
+y_4^{(m\ell)}\left( M_{\tilde{B}}^{(\ell k)} \right)^{-1}M_{M}^{(kp)}
\left( M_{\tilde{B}}^{(pq)} \right)^{-1}y_4^{(qn)}
\left( m_\Phi^2 \right)^{-1}
\xi_{\bf 7}'v_\phi^{2}\sin^2\theta_\phi,
\end{align}
where $M_{\tilde{B}}$ is the mass of $\tilde{B}$ defined in
Table~\ref{Tab:Matter_Content-GUT-fermion-G521'}, 
$m_X^2$ is the mass of $X$,
and $M_M$ is the Majorana mass given in Eq.~(\ref{Eq:Lagrangian-SU7}).
With one generation and assuming $M_{\tilde{B}}\simeq M_M$,
we obtain 
\begin{align}
y_{D\Delta}\simeq 
y_4^2\sin^2\theta_\phi
\frac{\xi_{\bf 7}'v_\phi^{2}}{ M_Mm_\Phi^2}.
\end{align}
Third, both $D^{(m)}$ and $N^{(m)}$ fermions also have the Dirac mass terms via the VEV of
$\Phi$.
Fourth, $N^{(m)}$ have the effective Majorana mass term via the VEV of 
$X$. The mass term of $N^{(m)}$ is generated by the VEV of $X$ and
mediated by $\tilde{B}$:
\begin{align}
 M_N^{(mn)}&=y_5^{(m\ell)}v_X
 \left( M_{\tilde{B}}^{(\ell k)} \right)^{-1}
 M_M^{(kp)}
 \left( M_{\tilde{B}}^{(pq)} \right)^{-1}
 y_5^{(qn)}v_X,
\end{align}
where 
\begin{align}
M_{\tilde{B}}^{(mn)}=M_M^{(mn)}+y_6^{(mn)}v_\Omega.
\end{align}
We assume that the mass of $\tilde{B}^{(m)}$ is of ${\cal O}(M_I)$.
The $G_{\rm SM}$ singlet fermion $\tilde{B}^{(m)}$ contributes to the
light neutrino mass through the Type-I seesaw mechanism
\cite{Minkowski:1977sc,Yanagida:1979as}.

Finally, we come to the consideration of the neutrino mass.  We first consider neutrino masses via
the Type-III seesaw mechanism
\cite{Foot:1988aq}. The $SU(2)_L$ triplet fermion
$\tilde{W}^{(m)}$ in Table~\ref{Tab:Matter_Content-GUT-fermion-G521'} required for the Type-III seesaw mechanism is
contained in $\Psi_{\bf 48}^{(m)}$. The masses of $\tilde{W}^{(m)}$,
$M_{\tilde{W}}^{(mn)}$, are given by 
\begin{align}
M_{\tilde{W}}^{(mn)}
=M_M^{(mn)}+y_6^{(mn)}v_\Omega,
\end{align}
which are assumed to be of ${\cal O}(M_I)$. 
The coupling constant matrix of the SM Higgs boson, neutrinos, and
$SU(2)_L$ triplet fermions is $y_4^{(mn)}$.
Thus, the effective mass matrix of the light neutrinos is given by
\begin{align}
M_\nu^{(mn)}=
\left( y_4^{(m\ell)}v\sin{\theta_h} \right)
\left(M_{\tilde{W}}^{(\ell k)}\right)^{-1}
M_M^{(kp)}
\left(M_{\tilde{W}}^{(pq)}\right)^{-1}
\left( y_4^{(qn)}v\sin{\theta_h} \right).
\label{Eq:Type-III-seesaw}
\end{align}
Next, the $G_{\rm SM}$ singlet fermions $N^{(m)}$ also contribute to the
light neutrino masses through the Type-I seesaw mechanism. 
The mass matrix of the right-handed neutrino $N^{(m)}$ is given by 
\begin{align}
 M_N^{(mn)}&=y_5^{(m\ell)}v_X
 \left( M_{\tilde{B}}^{(\ell k)} \right)^{-1}
 M_M^{(kp)}
 \left( M_{\tilde{B}}^{(pq)} \right)^{-1}
 y_5^{(qn)}v_X,
\end{align}
where 
\begin{align}
M_{\tilde{B}}^{(mn)}=M_M^{(mn)}+y_6^{(mn)}v_\Omega.
\end{align}
The Yukawa coupling constant matrix of the neutrino multiplets is given
by
\begin{align}
y_\nu^{(mn)} =y_2^{(mn)}\sin{\theta_h}.
\end{align}
Thus, the effective mass matrix of the light neutrino is given by
\begin{align}
M_\nu^{(mn)}=
\left( y_2^{(m\ell)}v\sin{\theta_h} \right)
\left(M_N^{(\ell k)}\right)^{-1}
M_M^{(kp)}
\left(M_N^{(pq)}\right)^{-1}
\left( y_2^{(qn)}v\sin{\theta_H} \right).
\label{Eq:Type-I-seesaw}
\end{align}
For the one generation case, from 
Eqs.~(\ref{Eq:Type-III-seesaw}) and (\ref{Eq:Type-I-seesaw}), the
contribution to the neutrino masses from Type-III and Type-I seesaw
mechanisms are given respectively by
\begin{align}
M_\nu^{\rm Type\mbox{-}III}\simeq 
\frac{v^2\sin^2{\theta_h} y_4^2}{M_M},\ \ 
M_\nu^{\rm Type\mbox{-}I}\simeq 
\frac{y_2^2v^2M_M^3\sin^2{\theta_h}}{y_5^4v_X^4}.
\end{align}
The coupling matrix $y_2^{(mn)}$ has the constraint from reproducing the mass matrices of down-type quarks and charged leptons.  On the other hand, the coupling matrices $y_4^{(mn)}$ and $y_5^{(mn)}$ have no restriction except for the neutrino masses.  Therefore, we can choose them to reproduce the current neutrino masses and mixing matrix.

\section{Gauge coupling unification}
\label{Sec:RGE}

To determine the $G_{\rm SM}\times SU(2)_D$ scale, i.e.,
the intermediate scale $M_I$, and the magnitudes of the associated gauge coupling constants at the scale, we 
discuss the RGEs for gauge coupling constants running from the
electroweak scale $M_{\rm EW}=M_Z$, the $SU(2)_D$ breaking scale $M_D$,
the intermediate scale $M_{I}$, to the unification scale $M_U$.
We must take $M_D>O(1)$\,TeV to avoid the constraint on ``4th-generation'' quark masses $m_{t',b'}\gtrsim1.5$~TeV from the LHC experiments \cite{ParticleDataGroup:2022pth}.

The RGE for the gauge coupling constant
$\alpha_i(\mu):= g_i^2(\mu)/4\pi$ at one-loop level is given in, e.g.,
Refs.~\cite{Slansky:1981yr,Yamatsu:2015gut} by 
\begin{align}
\frac{d}{d \log\mu}\alpha_i^{-1}(\mu)
=
-\frac{b_i}{2\pi},
\label{Eq:RGE-gauge-coupling-alpha}
\end{align}
where $i$ labels a specific gauge group $G$ and the beta function coefficient
\begin{align}
b_i=
-\frac{11}{3}\sum_{\rm Vector}T(R_V)
+\frac{2}{3}\sum_{\rm Weyl}T(R_F)
+\frac{1}{6}\sum_{\rm Real}T(R_S),
\label{Eq:beta-function-coeff-general}
\end{align}
where 'Vector', 'Weyl', and 'Real' stand for real vector, Weyl fermion, and
real scalar fields, respectively.
Since the vector bosons here are all gauge bosons, they belong to the
adjoint representation of the Lie group $G$ and $T(R_V)=C_2(G)$.
Here $C_2(G)$ is the quadratic Casimir invariant of the adjoint
representation of $G$, and  
$T(R_i)$ is the Dynkin index of the irreducible representation $R_{i}$ of
$G$.
Note that when the group $G$ is spontaneously broken into its
subgroup $G'$, it is more convenient to use the irreducible representations
of $G'$. (For the Dynkin index and the branching rules, see,
e.g., Refs.~\cite{Yamatsu:2015gut,McKay:1981} or
they can be calculated by using appropriate computer programs such as
Susyno~\cite{Fonseca:2011sy}, LieART~\cite{Feger:2012bs,Feger:2019tvk},
and GroupMath~\cite{Fonseca:2020vke}.
When the beta function coefficient $b_i$ is a constant at least 
from one energy scale $\mu_0$ to another $\mu_1$, we can solve the RGE in
Eq.~(\ref{Eq:RGE-gauge-coupling-alpha}) as 
\begin{align}
 \alpha_i^{-1}(\mu_1)=\alpha_i^{-1}(\mu_0)
 -\frac{b_i}{2\pi}
 \mbox{log}\left(\frac{\mu_1}{\mu_0}\right).
\end{align}
When the beta function coefficients depend on the energy scale $\mu$, we
have to use the different values of the corresponding beta function
coefficient for each scale $\mu$.

If we connect different gauge coupling constants at some scale,
we need to know the matching conditions at the scale.
We consider the relations between gauge groups $G_1\times G_2$
and their subgroup $G$ at the energy scale $\mu$, where 
the corresponding generators $T_1$, $T_2$ and $T$ are related by
\begin{align}
 T=p_1T_1+p_2T_2.
 \label{Eq:T1-T2_T}
\end{align}
The corresponding gauge couplings $g_1$, $g_2$, and $g$ satisfy the
following relation at $\mu$:
\begin{align}
 \frac{1}{g^2(\mu)}=
 \frac{p_1^2}{g_1^2(\mu)}+\frac{p_2^2}{g_2^2(\mu)},
\end{align}
or equivalently,
\begin{align}
 \alpha(\mu)^{-1}=
 p_1^2 \alpha_1(\mu)^{-1} + p_2^2 \alpha_2(\mu)^{-1}.
 \label{Eq:alhpa1-alpha2_alpha}
\end{align}
For more detail, see, e.g., Sec.~7.4 in Ref.~\cite{Mohapatra2002} at the
one-loop level.

In the following, we discuss whether it is possible to unify the gauge
coupling constants by solving the RGEs under the assumed symmetry
breaking pattern 
$SU(7)\to G_{521}'\to G_{\rm SM}\times SU(2)_D \to G_{\rm SM}$.
The field content is shown in 
Tables~\ref{Tab:Matter_Content-GUT-fermion-G521'},
\ref{Tab:Matter_Content-GUT-scalar-G521'}, and  
\ref{Tab:Matter_Content-GUT-gauge-G521'}, where the mass of 
the $SU(5)_{\rm CD}$ and $SU(2)_L$ adjoint fermions in ${\bf 48}$ of
$SU(7)$ is assumed to be of ${\cal O}(M_I)$.

We will examine the RGEs and matching conditions one by one below, as the
type of gauge symmetry changes at each scale in this calculation.
\begin{itemize}
 \item  
At $\mu=M_Z=91.1876\pm 0.0021$$\,$GeV, the input parameters for the three SM gauge
coupling constants are given in Ref.~\cite{ParticleDataGroup:2022pth}:
\begin{align}
\alpha_{3C}(M_Z)=0.1181 \pm 0.0011,\ \ 
\alpha_{2L}(M_Z)=\frac{\alpha_{\rm EM}(M_{Z})}{\sin^2\theta_W(M_Z)},\ \
\alpha_{1Y}(M_Z)=\frac{5\alpha_{\rm EM}(M_{Z})}{3\cos^2\theta_W(M_Z)},
\label{Eq:Input-1}
\end{align}
where the experimental values of the electromagnetic gauge coupling constant
$\alpha_{\rm EM}$ and the weak angle are given as
\begin{align}
\alpha_{\rm EM}^{-1}(M_{Z})=127.955\pm 0.010,\ \ 
\sin^2\theta_W(M_Z)=0.23122\pm 0.00003.
\label{Eq:Input-2}
\end{align}
 \item  
For $M_Z<\mu< M_D$, the gauge coupling constants of
$G_{\rm SM}=SU(3)_C\times SU(2)_L\times U(1)_Y$ evolve according to
\begin{align}
\alpha_{i}^{-1}(\mu)&=
\alpha_{i}^{-1}(M_Z)-\frac{b_{i}}{2\pi}\log\left(\frac{\mu}{M_Z}\right),
\ \ \ 
\left(
\begin{array}{c}
b_{3C}\\
b_{2L}\\
b_{1Y}\\
\end{array}
 \right)
=\left(
\begin{array}{c}
-7\\
-19/6\\
+41/10\\
\end{array}
 \right),
\label{Eq:RGE-SM}
\end{align}
where $i=3C,2L,1Y$ stand for 
$SU(3)_C$, $SU(2)_L$, $U(1)_Y$, respectively, 
$b_i$ are the beta function coefficients of $G_{\rm SM}$
for $M_Z<\mu< M_D$,
and we take the $SU(5)$ normalization for $U(1)_Y$.
 \item  
At $\mu=M_D$, by using Eq.~(\ref{Eq:RGE-SM}), the matching
conditions between $G_{\rm SM}$ and $G_{\rm SM}\times SU(2)_D$ are
\begin{align}
&\alpha_{3C;D}(M_D)=\alpha_{3C}(M_D),\ \
\alpha_{2L;D}(M_D)=\alpha_{2L}(M_D),\ \
\alpha_{1Y;D}(M_D)=\alpha_{1Y}(M_D),
\nonumber\\
&\alpha_{2D;D}(M_D)
=\mbox{[No need for matching]},
\label{Eq:Matching-MD}
\end{align}
where $\alpha_{3C;D}(\mu)$, $\alpha_{2L;D}(\mu)$, $\alpha_{1Y;D}(\mu)$,
$\alpha_{2D;D}(\mu)$ represent 
the gauge coupling constants of $SU(3)_C$, $SU(2)_L$, $U(1)_Y$, $SU(2)_D$,
respectively,
and $\alpha_{2D{;D}}(M_D)$ cannot be fixed only by using SM gauge
couplings without employing matching conditions from grand unification.
 \item  
For $M_D<\mu< M_I$, the RGEs of the gauge structure constants of
$G_{\rm SM}\times SU(2)_D$ are given by 
\begin{align}
\alpha_{i;D}^{-1}(\mu)&=
\alpha_{i;D}^{-1}(M_D)
-\frac{b_{i;D}}{2\pi}\log\left(\frac{\mu}{M_D}\right),
\ \ \ 
\left(
\begin{array}{c}
b_{3C;D}\\
b_{2L;D}\\
b_{1Y;D}\\
b_{2D;D}\\
\end{array}
 \right)
=\left(
\begin{array}{c}
-3\\
+5/6\\
+81/10\\
+7/6\\
\end{array}
 \right),
\label{Eq:RGE-SM+D}
\end{align}
where 
$i=3C,2L,1Y,2D$ 
stand for $SU(3)_C$, $SU(2)_L$, $U(1)_Y$, $SU(2)_D$, respectively,
and $b_{i;D}$ are the beta function coefficients of 
$G_{\rm SM}\times SU(2)_D$ in this energy regime.
 \item  
At $\mu= M_I$, 
the matching conditions between $G_{\rm SM}\times SU(2)_D$ from Eq.~(\ref{Eq:RGE-SM+D}) and
$G_{521}'=SU(5)_{CD}\times SU(2)_L\times U(1)_\alpha$
are
\begin{align}
\alpha_{CD;I}^{-1}(M_I)&=
 \alpha_{3C;D}^{-1}(M_I)
 =\alpha_{2D;D}^{-1}(M_I)
 ,\allowdisplaybreaks[1]\nonumber\\ 
\alpha_{2L;I}^{\prime-1}(M_I)&=
\alpha_{2L;D}^{-1}(M_I),\allowdisplaybreaks[1]\nonumber\\ 
\alpha_{1\alpha;I}^{-1}(M_I)&=
\frac{25}{21}\alpha_{1Y;D}^{-1}(M_I)
-\frac{4}{21}\alpha_{3C;D}^{-1}(M_I).
\end{align}
where $\alpha_{CD;I}(\mu)$, $\alpha_{2L;I}(\mu)$, 
$\alpha_{1\alpha;I}(\mu)$ represent 
the gauge structure constants of $SU(5)_{CD}$, $SU(2)_{2L}$, $U(1)_{1\alpha}$,
respectively,
 \item  
For $M_I<\mu< M_U$, the RGEs of the gauge
coupling constants of $G_{521}'$  are given by 
\begin{align}
\alpha_{i;I}^{-1}(\mu)&=
\alpha_{i;I}^{-1}(M_I)
-\frac{b_{i;I}}{2\pi}\log\left(\frac{\mu}{M_I}\right),\ \ \ 
\left(
\begin{array}{c}
b_{CD;I}\\
b_{2L;I}\\
b_{1\alpha;I}\\
\end{array}
 \right)
=\left(
\begin{array}{c}
+13/6\\
+19/3\\
+178/21\\
\end{array}
 \right),
\label{Eq:RGE-I}
\end{align}
where $i=CD,2L,1\alpha$ stand for $SU(5)_{CD}$, $SU(2)_L'$,
$U(1)_{\alpha}$, respectively.
 \item  
For $\mu=M_U$, the matching condition
between $G_{521}'$ from Eq.~(\ref{Eq:RGE-I}) and $SU(7)$ at $\mu=M_U$ is given by
\begin{align}
\alpha_{CD;I}^{-1}(M_U)=
\alpha_{2L;I}^{-1}(M_U)=
\alpha_{1\alpha;I}^{-1}(M_U)=:\alpha_U^{-1},
\label{Eq:Matching-condition-MU-G521'} 
\end{align}
where $\alpha_U$ stands for the gauge structure constant of $SU(7)$ at 
$\mu=M_U$.

\end{itemize}

\begin{figure}[tbh]
\begin{center}
\includegraphics[bb=0 0 481 348,height=5.5cm]{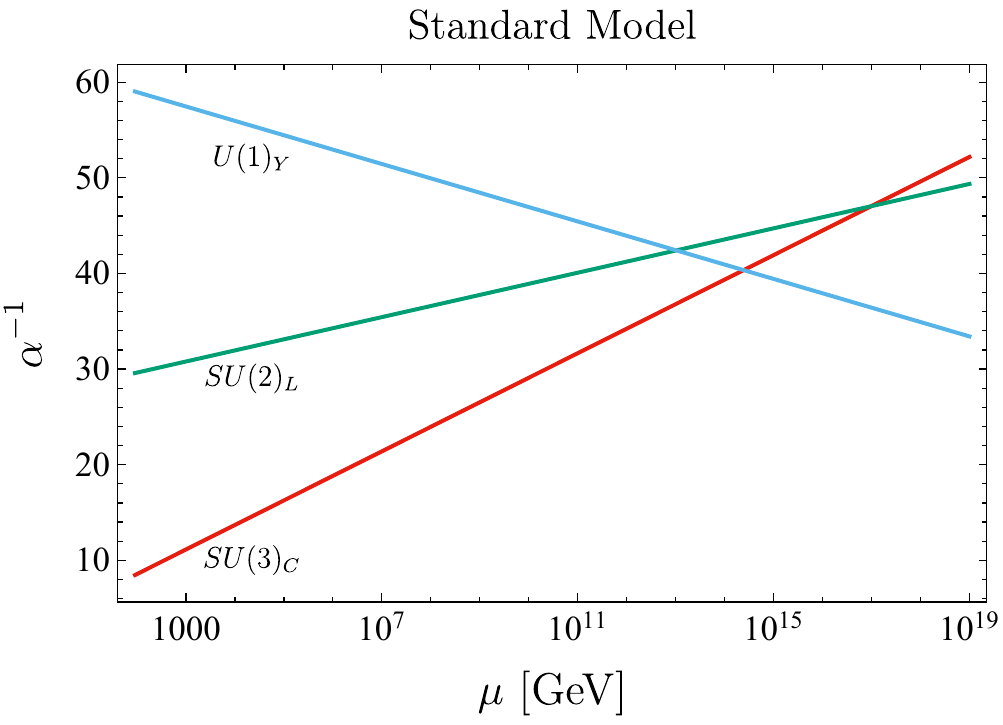}
\includegraphics[bb=0 0 480 339,height=5.5cm]{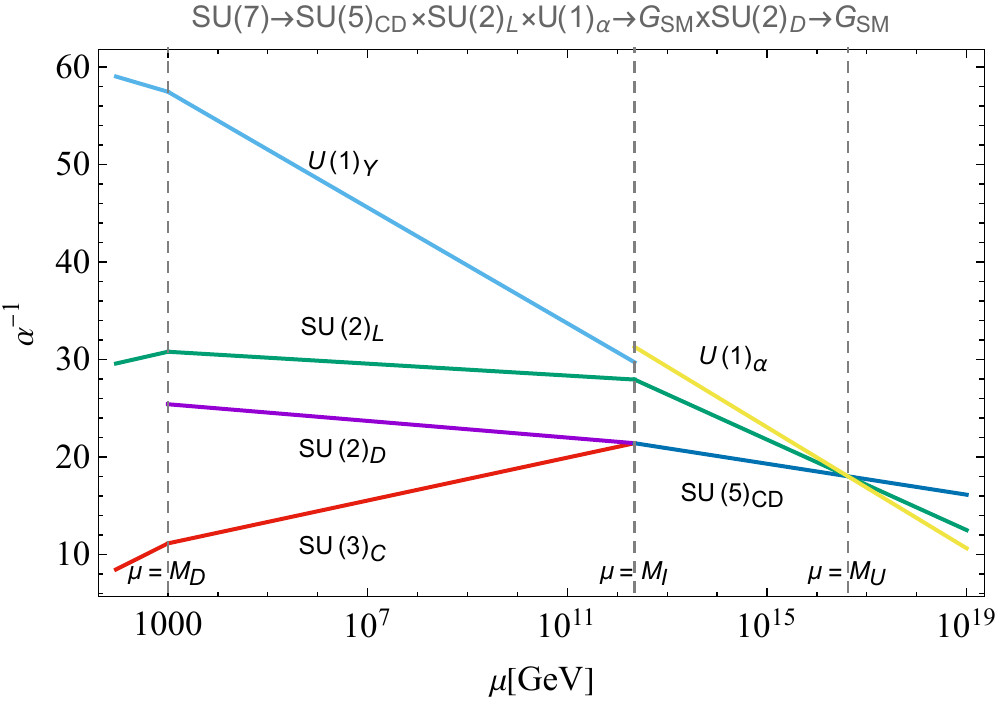}
\end{center}
\caption{\small The gauge coupling constants $\alpha_i$ vs the energy
 scale 
 $\mu$ for the SM (left) and the $SU(7)$ GUT model (right).
 The left plot shows the energy dependence of three gauge coupling
 constants of $SU(3)_C$,  $SU(2)_L$, and  $U(1)_Y$,
 $\alpha_{3C}$, $\alpha_{2L}$, and $\alpha_{1Y}$
 in the energy range of
 $\mu=[M_Z,M_H]$, where $M_H=10^{19}$$\,$GeV.
 The right plot shows 
 $\alpha_{3C}$, $\alpha_{2L}$, and $\alpha_{1Y}$
 in the energy ranges of $\mu=[M_Z,M_D]$;
 $\alpha_{3C}$, $\alpha_{2L}$, $\alpha_{1Y}$, and $\alpha_{2D}$
 in the energy range of $\mu=[M_D,M_I]$; and
 $\alpha_{CD}$, $\alpha_{2L}$,  $\alpha_{1\alpha}$ in the energy range of
 $\mu=[M_I,M_H]$
 \cite{ParticleDataGroup:2022pth}.
}
\label{Figure:RGE-gauge-coupling-G521'}
\end{figure}

From the matching condition in Eq.~(\ref{Eq:Matching-condition-MU-G521'}),
we can analytically solve the intermediate scale $M_I$
and unification scale $M_U$ as 
\begin{align}
 M_I&=M_D\ \mbox{exp}\left[\Xi\right],
\nonumber\\
 M_U&
 =
 M_I\ \mbox{exp}\left[
 \frac{A_0}{A_3}-\frac{A_1}{A_3}\log\left(\frac{M_D}{M_Z}\right)
 -\frac{A_2}{A_3}\Xi
 \right],
\nonumber\\
 \Xi&:=\frac{B_3 A_0-A_3 B_0}{B_3A_2-A_3B_2}+
 -\frac{B_3 A_1-A_3 B_1}{B_3A_2-A_3B_2}
 \log\left(\frac{M_D}{M_Z}\right),
\label{Eq:MI-MU-G521'}
\end{align}
where 
\begin{align}
 &A_0=\alpha_{3C}^{-1}(M_Z)-\alpha_{2L}^{-1}(M_Z),\
 A_1=\frac{b_{3C}-b_{2L}}{2\pi},\nonumber\\
 &A_2=\frac{b_{3C;D}-b_{2L;D}}{2\pi},\ 
 A_3=\frac{b_{CD;I}-b_{2L;I}}{2\pi},\nonumber\\
 &B_0=\frac{25}{21}\left(
 \alpha_{3C}^{-1}(M_Z)-\alpha_{1Y}^{-1}(M_Z)\right),\ 
 B_1=\frac{25}{21}\frac{b_{3C}-b_{1Y}}{2\pi},\nonumber\\
 &B_2=\frac{25}{21}\frac{b_{3C;D}-b_{1Y;D}}{2\pi},\ \
 B_3=\frac{b_{CD;I}-b_{1\alpha;I}}{2\pi}.
\end{align}
By solving Eq.~(\ref{Eq:MI-MU-G521'}), we find the values of the $M_I$
and $M_U$ as 
\begin{align}
 M_I\simeq 2.22\times 10^{12}~\mbox{GeV},\ \ 
 M_U\simeq 4.18\times 10^{16}~\mbox{GeV},
 \label{Eq:MI-MU-value-G521'}
\end{align}
where we have taken $M_D=1$~TeV.
The unified gauge coupling constant at $\mu=M_U$ is given by
\begin{align}
 \alpha_U^{-1}\simeq 18.02.
 \label{Eq:AlphaU-value-G521'}
\end{align}
The energy dependence of the gauge coupling constants $\alpha_i(\mu)$
in the $SU(7)$ GUT model is plotted in
Fig.~\ref{Figure:RGE-gauge-coupling-G521'}.
Table~\ref{Tab:Summary-Ng-Nl} shows $M_I$, $M_U$, and $\alpha_U^{-1}$ for different choices of $(N_g,N_\ell)$, where 
where $N_g$ and $N_\ell$ denote the numbers of the $SU(5)_{CD}$ adjoint and $SU(2)_{L}$ adjoint fermions of the
intermediate scale mass ${\cal O}(M_I)$, respectively. 
Our discussion so far assumes the case $(N_g,N_\ell)=(3,3)$.

\begin{table}[thb]
\begin{center}
\begin{tabular}{|c||c|c|c|}\hline
 \rowcolor[gray]{0.8}
 $(N_g,N_\ell)$ & $M_I/$GeV & $M_U/$GeV & $\alpha_U^{-1}$
 \\\hline\hline
 $(3,3)$& 2.22$\times10^{12}$ & 4.18$\times10^{16}$ &18.02 \\ \hline
 $(2,3)$& 6.89$\times10^{12}$ & 9.16$\times10^{14}$ &22.90 \\ \hline
 $(1,3)$& 1.00$\times10^{13}$ & 2.59$\times10^{14}$ &24.46 \\ \hline
 $(0,3)$& 1.20$\times10^{13}$ & 1.38$\times10^{14}$ &25.26 \\ \hline
 $(3,2)$& -- & -- & --\\ \hline
 $(2,2)$& 1.68$\times10^{12}$ & 1.54$\times10^{15}$ &22.54 \\ \hline
 $(1,2)$& 4.77$\times10^{12}$ & 2.63$\times10^{14}$ &24.65 \\ \hline
 $(0,2)$& 7.35$\times10^{12}$ & 1.26$\times10^{14}$ &25.52 \\ \hline
\end{tabular}
\begin{tabular}{|c||c|c|c|}\hline
 \rowcolor[gray]{0.8}
 $(N_g,N_\ell)$ & $M_I/$GeV & $M_U/$GeV & $\alpha_U^{-1}$
 \\\hline\hline
 $(3,1)$& -- & -- & --\\ \hline
 $(2,1)$& 6.43$\times10^{10}$ & 5.17$\times10^{15}$ &21.82 \\ \hline
 $(1,1)$& 1.45$\times10^{12}$ & 2.68$\times10^{14}$ &24.95 \\ \hline
 $(0,1)$& 3.69$\times10^{12}$ & 1.10$\times10^{14}$ &25.89 \\ \hline
 $(3,0)$& -- & -- & --\\ \hline
 $(2,0)$& -- & -- & --\\ \hline
 $(1,0)$& 1.56$\times10^{11}$ & 2.80$\times10^{14}$ &25.51 \\ \hline
 $(0,0)$& 1.32$\times10^{12}$ & 9.07$\times10^{13}$ &26.43 \\ \hline
\end{tabular}
 \caption{\small 
 Values of $M_I$, $M_U$, and $\alpha_U^{-1}$ for different numbers of $SU(5)_{CD}$ adjoint and
 $SU(2)_{L}$ adjoint fermions of the intermediate scale mass ${\cal O}(M_I)$.
 $N_g$ and $N_\ell$ denote the numbers of the $SU(5)_{CD}$ adjoint and
 $SU(2)_{L}$ adjoint fermions, respectively.
 The case with dashed entries (--) has no solution with 
 $M_U\leq 10^{19}$~GeV.
 }
\label{Tab:Summary-Ng-Nl}
\end{center}  
\end{table}

We comment on the proton decay. In the model, a proton can decay via the so-called leptoquark gauge bosons. The lifetime of proton mediated via these gauge bosons is roughly estimated as
\cite{Nath:2006ut,Mohapatra2002,ParticleDataGroup:2022pth},
\begin{align}
\tau\simeq\frac{M_U^4}{\alpha_U^2 m_p^5},
\label{Eq:proton-lifetime}
\end{align}
where $m_p$ is the proton
mass and the gauge boson masses are assumed to be $M_U$.
By using $M_U$ in Eq.~(\ref{Eq:MI-MU-value-G521'}) and $\alpha_U$ in
Eq.~(\ref{Eq:AlphaU-value-G521'}) for the case $(N_g,N_\ell)=(3,3)$, we
obtain 
\begin{align}
\tau\simeq 2.85\times 10^{37}\, \mbox{years}.
\end{align}
The current constraint is 
$\tau(p\to e^+\pi^0)>2.4\times 10^{34}$~years at $90\%$ CL
\cite{Super-Kamiokande:2020wjk}. 
We find that this proton decay constraint cannot be satisfied by all 
the other cases listed in Table~\ref{Tab:Summary-Ng-Nl}.

We have also performed the same analysis for the other symmetry-breaking
schemes, but there is no solution that satisfies the gauge coupling unification
and the proton decay constraint simultaneously.

\section{Dark matter (in)stability}
\label{Sec:DM}

We briefly review how the $U(1)_V$ symmetry is realized for a vacuum in the $G_{\rm SM}\times SU(2)_D$ pNGB DM
model.  By rewriting $\Phi(x)$ by $\Sigma(x):=(i\sigma_2\Phi^*(x),\Phi(x))$ 
in Eq.~(\ref{Eq:Potential-scalar}), the scalar potential is given as 
\begin{align}
 {\cal V}(H,{\Sigma},\Delta)
 &=
 -\mu_H^2H^\dag H
 -\frac{\mu_\Phi^2}{2}
 \mbox{Tr}\left({\Sigma}^\dag{\Sigma}\right)
 -\frac{1}{2}\mu_{\Delta}^2\mbox{Tr}\left(\Delta^2\right)
 \nonumber\\
 &\hspace{1em}
 -
 \sqrt{2}
 \kappa_1
 \mbox{Tr}
 \left(
 \sigma_1
 {\Sigma}^\dag \Delta{\Sigma}
 \right)
 -
 \sqrt{2}
 \kappa_2
 \mbox{Tr}
 \left(
 \sigma_2
 {\Sigma}^\dag \Delta{\Sigma}
 \right)
 -\sqrt{2}\kappa_3
 \mbox{Tr}
 \left(
 \sigma_3
 {\Sigma}^\dag \Delta{\Sigma}
 \right)
 \nonumber\\
 &\hspace{1em}
 +\lambda_{H}\left(H^\dag H\right)^2
 +\frac{\lambda_\Phi}{4}
 \left(\mbox{Tr}\left({\Sigma}^\dag{\Sigma}\right)\right)^2
 +\frac{1}{4}\lambda_{\Delta}\mbox{Tr}\left(\Delta^2\right)^2
 \nonumber\\
 &\hspace{1em}
 +\frac{1}{2}\lambda_{H\Phi}
 \left(H^\dag H\right)
 \mbox{Tr}\left({\Sigma}^\dag{\Sigma}\right)
 +\frac{1}{2}\lambda_{H\Delta}\left(H^\dag H\right)
 \mbox{Tr}\left(\Delta^2\right)
 +\frac{1}{2}\lambda_{\Phi\Delta}
 \mbox{Tr}\left({\Sigma}^\dag{\Sigma}\right)
  \mbox{Tr}\left(\Delta^2\right). 
\label{Eq:Potential-scalar-check}
\end{align}
When we neglect the fermionic sector, $\kappa_1=\kappa_2=0$ can be 
chosen without loss of generality by the redefinition of $\Sigma$. 
In this basis, for $\mu_\Phi ^2>0$ and $\mu_\Delta^2>0$, 
the global minima of the vacuum are given by 
\begin{align}
 \langle\Delta\rangle&=\frac{1}{\sqrt{2}}\left(
 \begin{array}{cc}
  v_{\Delta}^{}&0\\
  0&-v_{\Delta}^{}\\
 \end{array}
 \right),\ \ \
 \langle{\Sigma}\rangle=
 \frac{1}{\sqrt{2}}
 \left(
 \begin{array}{cc}
  v_{\Phi}^{}&0\\
  0&v_{\Phi}^{}\\
 \end{array}
 \right). 
\label{Eq:VEV-full}
\end{align}
See Ref.~\cite{Otsuka:2022zdy} for the details of the potential analysis.

The existence of the $U(1)_V$ global symmetry in the scalar potential is apparent if its transformation is given by $U_V=\mbox{exp}\left[i\theta_V^{}\frac{\sigma_3}{2}\right]$ because $[U_V, \langle \Delta \rangle]=0$ and $[U_V, \langle \Sigma \rangle]=0$.
Under the $U(1)_V$, the scalar fields transform as 
\begin{align}
\Delta \to e^{i\theta_V^{}\frac{\sigma_3}2}\, \Delta\, e^{-i\theta_V^{}\frac{\sigma_3}2}, 
\quad 
\Sigma \to e^{i\theta_V^{}\frac{\sigma_3}2}\, \Sigma\, e^{-i\theta_V^{}\frac{\sigma_3}2}. 
\end{align}
We then read off $U(1)_V$ charges of the fields 
in the broken phase as 
\begin{align}
\label{eq:dm-stability_su2Dmultiplet}
\Delta&=\left(
 \begin{array}{cc}
\eta_0/\sqrt2&\eta_{+}\\
 \eta_{-}&-\eta_0/\sqrt2\\
 \end{array}
 \right),\ \ \
\Sigma=
 \left(
 \begin{array}{cc}
  \phi_0^*&\phi_+\\
  -\phi_-&\phi_0\\
 \end{array}
 \right),
\end{align}
where the $U(1)_V$ charges are indicated by the subscript.

Since the vacuum preserves exact $U(1)_V$ symmetry, 
the lightest particle charged under the $U(1)_V$ would be stable. 
In addition, the pNGB DM model satisfies the thermal relic abundance of 
DM \cite{Aghanim:2018eyx} without conflicting with the constraint from the 
direct detection \cite{LZ:2022lsv}. 
The same results hold in the $SU(7)$ pNGB DM model with an
appropriate choice of parameters. 

Let us comment on the inconsistency of $U(1)_V$ with the $SU(7)$ pNGB DM model 
by focusing on the down-type quark sector, 
\begin{align}
 {\cal L}_d
 &=
 -\sum_{m=1}^3\sum_{n=1}^9
\tilde{d}^{(m)}d^{c(n)} \Bigg\{ y_{Dd}^{(mn)} 
 \left(
 \begin{array}{cc}
  \phi_0^*\\
  -\phi_-\\
 \end{array}
 \right)
+  y_{Dd}^{\prime(mn)}  
 \left(
 \begin{array}{c}
\phi_+\\
\phi_0\\
 \end{array}
 \right)
\Bigg\}
 +\mbox{H.c.}
\end{align}
From this expression, it is clear that the $U(1)_V$ transformations for $\phi_0$ and $\phi_\pm$ conflict each other.  Thus, $U(1)_V$ is broken explicitly by the Yukawa interaction.  We note that if one of the two Yukawa matrices, $y_{Dd}^{(mn)}$ and $y_{Dd}^{\prime(mn)}$, vanishes, $U(1)_V$ is exact with an appropriate choice of charges for the upper and lower components of $\tilde{d}$.  The leptonic Yukawa interactions including the Majorana mass terms also induce similar contradictions with the $U(1)_V$ symmetry.  As a result, non-zero $\kappa_1$ and $\kappa_2$ are generated through these Yukawa interactions, even if we set $\kappa_1=\kappa_2=0$ at the tree level in the $SU(7)$ pNGB DM model. This is also consistent with the $U(1)_V$ 
transformations of $\kappa_1$ and $\kappa_2$ terms, 
\begin{align}
\mbox{Tr}
 \left(
 (\kappa_1\sigma_1+ \kappa_2\sigma_2)
 {\Sigma}^\dag \Delta{\Sigma}
 \right)
\to 
\mbox{Tr}
 \left(
 e^{i\theta_V \sigma_3}
 (\kappa_1\sigma_1+ \kappa_2\sigma_2)
 {\Sigma}^\dag \Delta{\Sigma}
 \right). 
\end{align}
Therefore, a single type of Yukawa interaction $\{ y_{Dd}^{(mn)}=0$ or $y_{Dd}^{\prime(mn)}=0\}$ with $\kappa_1 = \kappa_2 = 0$ is the condition to preserve the $U(1)_V$ symmetry.\footnote{
The conditions $\kappa_1=\kappa_2=0$ can be realized if we impose the global symmetries $SU(2)_{\Phi}$ and $SU(2)_\Delta$, where $SU(2)_{\Phi}$ and $SU(2)_\Delta$ are global symmetries related with $\Phi$ and $\Delta$, respectively. These symmetries may be originated from the global symmetries $SU(7)_{\Phi_{\bf 7}}$ and $SU(7)_{\Phi_{\bf 48}}$ when considered in the original $SU(7)$ Lagrangian, where $SU(7)_{\Phi_{\bf 7}}$ and $SU(7)_{\Phi_{\bf 48}}$ are global symmetries related with $\Phi_{\bf 7}$ and $\Phi_{\bf 48}$, respectively.}

We comment on how we can realize the Lagrangian such that the $U(1)_V$ symmetry is preserved by extending the model. 
One may introduce a new $\mathbb{Z}_2$ symmetry in our model. We assign the new $\mathbb{Z}_2$ odd parity to $\Psi_{\bf 48}$ and $\Phi_{\bf 48}$, and even parity to all the other fields. The $\mathbb{Z}_2$ symmetry does not allow the Yukawa interaction terms with the coefficients $y_4^{(mn)}$, $y_5^{(mn)}$, and $y_6^{(mn)}$ in Eq.~(2.1). Therefore, the primed Yukawa interactions are not generated from higher-order operators.  
At the same time, $\kappa_1$ and $\kappa_2$ are also forbidden since $\xi_{\bf y}'$ $({\bf y}={\bf 7,21,35,48})$ in Eq.~(2.4), and $\lambda_{x}$ $(x=1,2,3,4)$ in Eq.~(2.5) are vanishing due to the $\mathbb{Z}_2$ symmetry. 
As a result, DM stability is guaranteed. 
On the other hand, one problem arises with this new $\mathbb{Z}_2$ symmetry: the Majorana mass term of the right-handed neutrinos $N^{(m)}$ is not generated. Therefore, it is not possible to produce tiny neutrino masses via the seesaw mechanism. 
The problem can be solved by introducing an additional scalar field $\Phi_{\bf 196}$ in ${\bf 196}$ of $SU(7)$.
With $\Phi_{\bf 196}$ there are an additional Yukawa interaction term 
$y_7^{(mn)}\Phi_{\bf 196}^\dag\left(\Psi_{\bf 21}^{(m)}\Psi_{\bf 21}^{(n)}\right)_{\bf 196}$ and a cubic scalar interaction term $\xi_3\Phi_{\bf 21}\Phi_{\bf 21}\Phi_{\bf 196}^\dag$. 
Through these interactions, the Majorana mass for the right-handed neutrinos in Eq.~(2.28) attains an additional contribution. 
Even when we introduce the $\mathbb{Z}_2$ symmetry in the $SU(7)$ invariant Lagrangian and the $U(1)_V$ symmetry is preserved by the vacuum, the Majorana mass term of $N^{(m)}$ can still be generated.

\section{Predictions of the SU(7) GUT pNGB DM model}
\label{Sec:Predictions}

In this section, we show the predictions of our model.  
We assume that the stability of DM is guaranteed by fine-tuning the coupling constants or by the above-mentioned extension of the model.
A remarkable point of the SU(7) GUT pNGB DM model is that the magnitudes of the gauge coupling constants are determined by the requirement of the gauge coupling unification.  As a result, we will show below that the DM mass is predicted to be at the weak scale in order to keep the pNGB as a DM candidate.

\begin{figure}[htb]
\begin{center}
\includegraphics[bb=0 0 670 580,height=10cm]{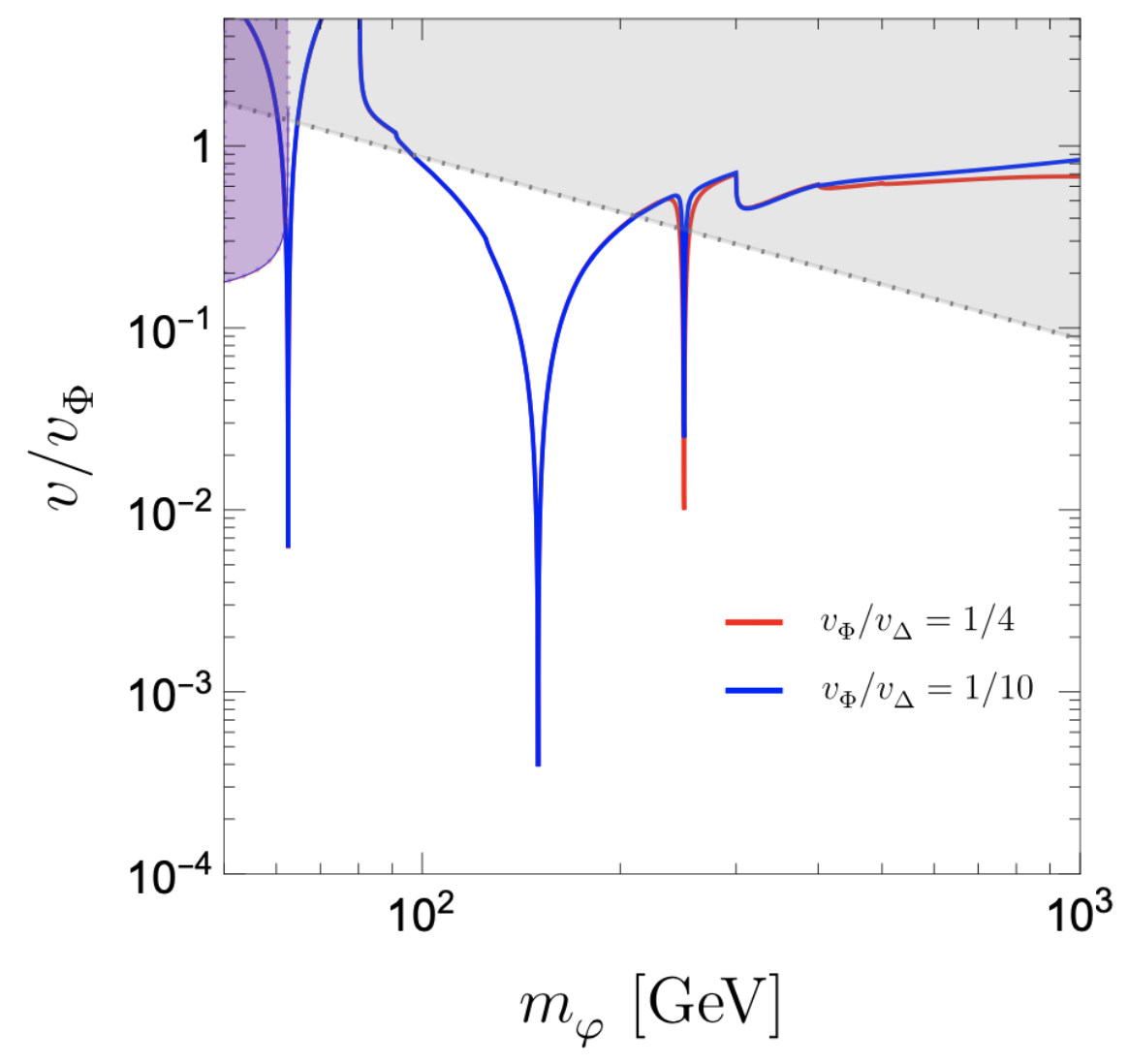}
 \caption{\small 
Current status of constraints on our GUT pNGB DM model.
Solid curves express parameter contours reproducing the correct dark matter energy density 
$\Omega h^2 =  0.12$.
 The gray-shaded region bounded by a dotted line is the region where the $SU(2)_D$ neutral gauge boson can also become a DM candidate, which is not considered in this analysis.
 The purple-shaded region is excluded by Higgs invisible decay constraints~\cite{ATLAS:2020kdi}.
 The constraint from the direct detection experiment is much above the bound of the gray-shaded region.
 }
\label{Fig:dd_vs_relic}
\end{center}
\end{figure}

The model parameters are chosen to be the same as in Ref.\cite{Otsuka:2022zdy} for comparison.  We take the mass of the second lightest scalar field $h_2$ as $m_{h_2}= 300$~GeV and the mass of the third lightest scalar field $h_3$ as $m_{h_3}= 500$~GeV.  Note that $h_2$, $h_3$, and the observed Higgs boson $h_1$ can be expressed as linear combinations of $\phi_0$ and $\eta_0$ in Eq.~(\ref{eq:dm-stability_su2Dmultiplet}) and the neutral CP-even scalar field $h_0$ coming from the $SU(2)_L$ doublet:
\begin{align}
\left(  \begin{array}{c}
         h_1
         \\
         h_2
         \\
         h_3
         \end{array}
\right)
=
\left(  \begin{array}{ccc}
         1 & 0 & 0
         \\
         0 & \cos\alpha_x & \sin\alpha_x 
         \\
         0 & -\sin\alpha_x & \cos\alpha_x 
         \end{array}
\right)  
\left(  \begin{array}{ccc}
         \cos\alpha_y & 0 &  \sin\alpha_y
         \\
         0 & 1 & 0 
         \\
         -\sin\alpha_y & 0 &  \cos\alpha_y
         \end{array}
\right)  
\left(  \begin{array}{ccc}
         \cos\alpha_z & \sin\alpha_z & 0
         \\
         -\sin\alpha_z & \cos\alpha_z & 0
         \\
         0 & 0 & 1 
         \end{array}
\right)         
\left(  \begin{array}{c}
         h_0
         \\
         \phi_0
         \\
         \eta_0
         \end{array}
\right).
\end{align}
In the following, the mixing angles among the scalar fields are taken to be $(\sin\alpha_x,\,\sin\alpha_y,\,\sin\alpha_z)=(0.06,0.05,0.1)$.  Figure~\ref{Fig:dd_vs_relic} shows the various constraints and the parameter regions that reproduce the correct dark matter relic abundance for different choices of the ratio of the VEV of the $SU(2)_D$ doublet scalar $v_\Phi$ to the VEV of the triplet scalar $v_\Delta$: $v_\Phi/v_\Delta=1/4,~1/10$.  The solid curves express the parameter contours that reproduce the DM energy density $\Omega h^2 = 0.12$.  The purple-shaded region is excluded by Higgs invisible decay constraints~\cite{ATLAS:2020kdi}.  The Higgs invisible decay width in our model shows $v_\Phi/v_\Delta$-dependence only through sub-leading terms and, therefore, the corresponding excluded region is common for $v_\Phi/v_\Delta = 1/4$ and $1/10$.

Depending on the mass spectrum of the $SU(2)_D$ gauge fields, the neutral component of the $SU(2)_D$ gauge boson can also be stabilized and be a dark matter candidate in addition to the pNGB $\varphi_\pm$.  Requiring the pNGB $\varphi_\pm$ as a single component DM in our universe, the gray-shaded region, where the mass of the neutral component of the $SU(2)_D$ gauge boson is less than twice the DM mass, is excluded.  Note that in our SU(7) GUT pNGB DM model, $g_D$ is uniquely determined by the condition for the gauge coupling unification, in contrast to Ref.~\cite{Otsuka:2022zdy} where the $SU(2)_D$ gauge coupling constant $g_D$ is a free parameter.  In this article, we use $g_D(\mu = M_D)=0.7033$ as an input at $M_D\simeq 1$~TeV.  Note that the gray shaded exclusion in this work becomes more stringent than that in Ref.~\cite{Otsuka:2022zdy}, where $g_D$ takes the maximum value allowed by unitarity.  Note that the constraint from the direct detection experiment is much above the bound of the gray-shaded region.

\section{Summary and discussions}
\label{Sec:Summary}

In this work, we have proposed a GUT pNGB DM model based on the $SU(7)$ gauge symmetry that
includes the SM gauge symmetry $G_{\rm SM}$ and the dark gauge symmetry 
$SU(2)_D$. The unification of SM fermions and dark sector fermions is
partially realized due to the GUT symmetry. 
We have examined all the symmetry-breaking patterns that preserve $G_{\rm SM}\times
SU(2)_D$ for $SU(7)$ GUT gauge symmetries.  In particular, we have found that
the phenomenologically successful cases are the following symmetry-breaking patterns.  First, the GUT gauge symmetry $SU(7)$ is spontaneously broken to $SU(5)_{CD}\times SU(2)_L\times U(1)_\alpha$
gauge symmetry at the GUT scale via the nonvanishing VEV of $SU(7)$
adjoint scalar field. The symmetry is broken to the SM gauge symmetry
$G_{\rm SM}$ and the dark gauge symmetry  $SU(2)_D$ at an intermediate
scale.  Furthermore, the $SU(2)_D$ symmetry is broken by the VEVs of the
$SU(2)_D$ doublet and triplet scalar fields at the TeV scale.
We have investigated the RG evolution of the gauge coupling constants for
the possible mass spectra of $SU(7)$ adjoint fermions.
The only pattern that satisfies the constraint from proton decay is
when there are three copies of the $SU(3)_C$ and $SU(2)_L$ adjoint
fermions with masses at the intermediate scale $M_I$. 
In particular, we find that the gauge coupling
unification and the proton decay constraints can be satisfied only by the
symmetry-breaking scheme of $SU(7)\to SU(5)_{CD}\times SU(2)_L\times U(1)_\alpha$.
This is because in the other cases, the unification scale $M_U$ is too
low to satisfy the constraint from proton decay.
At the same time, the intermediate scale $M_I\simeq 2.2\times
10^{12}$~GeV in this case is suitable for producing the observed tiny
neutrino masses through the type-I and type-III seesaw mechanisms
using the $G_{\rm SM}$ singlet fermions and $SU(2)_L$ triplet fermions.
The main contribution to the neutrino mass depends on the parameters of the model. 
In the pNGB DM model based on $G_{\rm SM}\times SU(2)_D$, the $U(1)_V$
custodial symmetry guarantees DM stability. On the other hand, in the $SU(7)$ pNGB DM model, this global symmetry is explicitly broken by the Yukawa interaction and the effective Majorana mass terms.
In the scalar sector, the cubic coupling constants of the $SU(2)_D$ doublet and triplet scalar fields are the order parameters of the $U(1)_V$ symmetry breaking.
In this model, $U(1)_V$ symmetry is essential for DM stability, and we need to tune Yukawa coupling constants and cubic scalar couplings such that $U(1)_V$ symmetry is satisfied at high accuracy.
We find that the allowed DM mass region is largely reduced since the gauge coupling constant of $SU(2)_D$, which was freely chosen in the previous pNGB DM model, is now fixed by the condition of the gauge coupling unification.

In the $SU(7)$ pNGB DM model, the gauge coupling constant of $SU(2)_D$,
which can be freely chosen in the $SU(2)_D$ pNGB DM model, is determined
by the requirement of gauge coupling unification.
From the discussion in Sec.~\ref{Sec:RGE}, the gauge coupling constant
of $SU(2)_D$ is ${\cal O}(1)$, comparable to the gauge coupling constant of $SU(2)_L$.
The scale of VEVs for breaking the $SU(2)_D$ symmetry is assumed to be $M_D$.
Thus, the mass of the $SU(2)_D$ gauge boson is of ${\cal O}(M_D)$.
The $SU(2)_D$ gauge boson does not directly couple to the SM fermions 
when we ignore the generational mixings of the down-type quarks and the
leptons in ${\bf 2}$ and ${\bf 1}$ of $SU(2)_D$ due to the Yukawa
coupling terms.

Since the information from the DM alone is not sufficient to discriminate this model from others, one must resort to other complementary experiments.
We therefore comment on the possibilities of exploring the signals of the
$SU(7)$ GUT pNGB DM model in collider experiments.
In the model, there are ``4th-generation'' down-type quarks and
charged leptons at the $SU(2)_D$ breaking scale $M_D$.
Such new particles will be further searched at the LHC or at future colliders beyond
the LHC energy scale, such as the FCC-hh\cite{FCC:2018vvp}.
A high-energy lepton collider, such as muon colliders\cite{Delahaye:2019omf}, may also be realized in the future.

\section*{Acknowledgments}

This work was supported in part by the National Science and Technology Council of Taiwan under Grant Nos.~NSTC-111-2112-M-002-018-MY3 (C.W.C.) and NSTC-111-2811-M-002-047-MY2 (N.Y.), by JSPS Grant-in-Aid for Scientific Research KAKENHI Grant No.~JP22K03620, JP18H05543 (K.T.), by the Guangdong Major Project of Basic and Applied Basic Research No.~2020B0301030008 (Y.U.), and by National Natural Science Foundation of China Grant No.~NSFC-12347112 (Y.U.).


\bibliographystyle{utphys}
\bibliography{../../arxiv/reference}

\providecommand{\href}[2]{#2}\begingroup\raggedright\begin{thebibliography}{10}

\bibitem{Georgi:1974sy}
H.~Georgi and S.~L. Glashow, ``{Unity of All Elementary Particle Forces},''
\href{http://dx.doi.org/10.1103/PhysRevLett.32.438}{{ Phys. Rev. Lett.}
  {\bfseries 32} (1974) 438--441}.

\bibitem{Slansky:1981yr}
R.~Slansky, ``{Group Theory for Unified Model Building},''
\href{http://dx.doi.org/10.1016/0370-1573(81)90092-2}{{ Phys. Rept.} {\bfseries
  79} (1981) 1--128}.

\bibitem{Yamatsu:2015gut}
N.~Yamatsu, ``{Finite-Dimensional Lie Algebras and Their Representations for
  Unified Model Building},''
\href{http://arxiv.org/abs/1511.08771}{{\ttfamily arXiv:1511.08771 [hep-ph]}}.

\bibitem{Inoue:1977qd}
K.~Inoue, A.~Kakuto, and Y.~Nakano, ``{Unification of the Lepton-Quark World by
  the Gauge Group SU(6)},''
\href{http://dx.doi.org/10.1143/PTP.58.630}{{ Prog.Theor.Phys.} {\bfseries 58}
  (1977) 630}.

\bibitem{Fritzsch:1974nn}
H.~Fritzsch and P.~Minkowski, ``{Unified Interactions of Leptons and
  Hadrons},''
\href{http://dx.doi.org/10.1016/0003-4916(75)90211-0}{{ Ann. Phys.} {\bfseries
  93} (1975) 193--266}.

\bibitem{Ida:1980ea}
M.~Ida, Y.~Kayama, and T.~Kitazoe, ``{Inclusion of Generations in SO(14)},''
\href{http://dx.doi.org/10.1143/PTP.64.1745}{{ Prog. Theor. Phys.} {\bfseries
  64} (1980) 1745}.

\bibitem{Fujimoto:1981bv}
Y.~Fujimoto, ``{SO(18) Unification},''
\href{http://dx.doi.org/10.1103/PhysRevD.26.3183}{{ Phys. Rev.} {\bfseries D26}
  (1982) 3183}.

\bibitem{Gursey:1975ki}
F.~Gursey, P.~Ramond, and P.~Sikivie, ``{A Universal Gauge Theory Model Based
  on $E_6$},''
\href{http://dx.doi.org/10.1016/0370-2693(76)90417-2}{{ Phys. Lett.} {\bfseries
  B60} (1976) 177}.

\bibitem{Kawamura:1999nj}
Y.~Kawamura, ``{Gauge Symmetry Breaking from Extra Space $S^1/Z_2$},''
  \href{http://dx.doi.org/10.1143/PTP.103.613}{{ Prog. Theor. Phys.} {\bfseries
  103} (2000) 613--619},
\href{http://arxiv.org/abs/hep-ph/9902423}{{\ttfamily arXiv:hep-ph/9902423
  [hep-ph]}}.

\bibitem{Hall:2001pg}
L.~J. Hall and Y.~Nomura, ``{Gauge Unification in Higher Dimensions},''
  \href{http://dx.doi.org/10.1103/PhysRevD.64.055003}{{ Phys.Rev.} {\bfseries
  D64} (2001) 055003},
\href{http://arxiv.org/abs/hep-ph/0103125}{{\ttfamily arXiv:hep-ph/0103125
  [hep-ph]}}.

\bibitem{Lim:2007jv}
C.~S. Lim and N.~Maru, ``{Towards a Realistic Grand Gauge-Higgs Unification},''
  \href{http://dx.doi.org/10.1016/j.physletb.2007.07.053}{{ Phys.Lett.}
  {\bfseries B653} (2007) 320--324},
\href{http://arxiv.org/abs/0706.1397}{{\ttfamily arXiv:0706.1397 [hep-ph]}}.

\bibitem{Kojima:2011ad}
K.~Kojima, K.~Takenaga, and T.~Yamashita, ``{Grand Gauge-Higgs Unification},''
  \href{http://dx.doi.org/10.1103/PhysRevD.84.051701}{{ Phys. Rev.} {\bfseries
  D84} (2011) 051701},
\href{http://arxiv.org/abs/1103.1234}{{\ttfamily arXiv:1103.1234 [hep-ph]}}.

\bibitem{Yamatsu:2017sgu}
N.~Yamatsu, ``{Special Grand Unification},''
  \href{http://dx.doi.org/10.1093/ptep/ptx088}{{ Prog. Theor. Exp. Phys.}
  {\bfseries 2017} no.~6, (2017) 061B01},
\href{http://arxiv.org/abs/1704.08827}{{\ttfamily arXiv:1704.08827 [hep-ph]}}.

\bibitem{Yamatsu:2018fsg}
N.~Yamatsu, ``{Family Unification in Special Grand Unification},''
  \href{http://dx.doi.org/10.1093/ptep/pty100}{{ Prog. Theor. Exp. Phys.}
  {\bfseries 2018} no.~9, (2018) 091B01},
\href{http://arxiv.org/abs/1807.10855}{{\ttfamily arXiv:1807.10855 [hep-ph]}}.

\bibitem{Nomura:2008sx}
T.~Nomura and J.~Sato, ``{Standard(-like) Model from an $SO(12)$ Grand Unified
  Theory in Six-Dimensions with $S_2$ Extra-Space},''
  \href{http://dx.doi.org/10.1016/j.nuclphysb.2008.11.017}{{ Nucl. Phys. B}
  {\bfseries 811} (2009) 109--122},
  \href{http://arxiv.org/abs/0810.0898}{{\ttfamily arXiv:0810.0898 [hep-ph]}}.

\bibitem{Hosotani:2015hoa}
Y.~Hosotani and N.~Yamatsu, ``{Gauge-Higgs Grand Unification},''
  \href{http://dx.doi.org/10.1093/ptep/ptv153}{{ Prog. Theor. Exp. Phys.}
  {\bfseries 2015} (2015) 111B01},
\href{http://arxiv.org/abs/1504.03817}{{\ttfamily arXiv:1504.03817 [hep-ph]}}.

\bibitem{Yamatsu:2017ssg}
N.~Yamatsu, ``{String-Inspired Special Grand Unification},''
  \href{http://dx.doi.org/10.1093/ptep/ptx135}{{ Prog. Theor. Exp. Phys.}
  {\bfseries 2017} no.~10, (2017) 101B01},
\href{http://arxiv.org/abs/1708.02078}{{\ttfamily arXiv:1708.02078 [hep-ph]}}.

\bibitem{Super-Kamiokande:2017gev}
{\bfseries Super-Kamiokande} Collaboration, K.~Abe {et al.}, ``{Search for
  Nucleon Decay into Charged Antilepton Plus Meson in 0.316 megaton$\cdot$years
  Exposure of the Super-Kamiokande Water Cherenkov Detector},''
  \href{http://dx.doi.org/10.1103/PhysRevD.96.012003}{{ Phys. Rev. D}
  {\bfseries 96} no.~1, (2017) 012003},
  \href{http://arxiv.org/abs/1705.07221}{{\ttfamily arXiv:1705.07221
  [hep-ex]}}.

\bibitem{Super-Kamiokande:2020tor}
{\bfseries Super-Kamiokande} Collaboration, M.~Tanaka {et al.}, ``{Search for
  Proton Decay into Three Charged Leptons in 0.37 Megaton-years Exposure of the
  Super-Kamiokande},'' \href{http://dx.doi.org/10.1103/PhysRevD.101.052011}{{
  Phys. Rev. D} {\bfseries 101} no.~5, (2020) 052011},
  \href{http://arxiv.org/abs/2001.08011}{{\ttfamily arXiv:2001.08011
  [hep-ex]}}.

\bibitem{Super-Kamiokande:2020wjk}
{\bfseries Super-Kamiokande} Collaboration, A.~Takenaka {et al.}, ``{Search for
  Proton Decay via $p\to e^+\pi^0$ and $p\to \mu^+\pi^0$ with an Enlarged
  Fiducial Volume in Super-Kamiokande I-IV},''
  \href{http://dx.doi.org/10.1103/PhysRevD.102.112011}{{ Phys. Rev. D}
  {\bfseries 102} no.~11, (2020) 112011},
  \href{http://arxiv.org/abs/2010.16098}{{\ttfamily arXiv:2010.16098
  [hep-ex]}}.

\bibitem{Super-Kamiokande:2022egr}
{\bfseries Super-Kamiokande} Collaboration, R.~Matsumoto {et al.}, ``{Search
  for Proton Decay via $p\rightarrow \mu^+K^0$ in 0.37 Megaton-years Exposure
  of Super-Kamiokande},''
  \href{http://dx.doi.org/10.1103/PhysRevD.106.072003}{{ Phys. Rev. D}
  {\bfseries 106} no.~7, (2022) 072003},
  \href{http://arxiv.org/abs/2208.13188}{{\ttfamily arXiv:2208.13188
  [hep-ex]}}.

\bibitem{JUNO:2022qgr}
{\bfseries JUNO} Collaboration, A.~Abusleme {et al.}, ``{JUNO Sensitivity on
  Proton Decay $p\to \bar\nu K^+$ Searches},''
  \href{http://arxiv.org/abs/2212.08502}{{\ttfamily arXiv:2212.08502
  [hep-ex]}}.

\bibitem{Hyper-Kamiokande:2018ofw}
{\bfseries Hyper-Kamiokande} Collaboration, K.~Abe {et al.},
  ``{Hyper-Kamiokande Design Report},''
  \href{http://arxiv.org/abs/1805.04163}{{\ttfamily arXiv:1805.04163
  [physics.ins-det]}}.

\bibitem{Heeck:2019kgr}
J.~Heeck and V.~Takhistov, ``{Inclusive Nucleon Decay Searches as a Frontier of
  Baryon Number Violation},''
  \href{http://dx.doi.org/10.1103/PhysRevD.101.015005}{{ Phys. Rev. D}
  {\bfseries 101} no.~1, (2020) 015005},
  \href{http://arxiv.org/abs/1910.07647}{{\ttfamily arXiv:1910.07647
  [hep-ph]}}.

\bibitem{Minkowski:1977sc}
P.~Minkowski, ``{$\mu \to e\gamma$ at a Rate of One Out of $10^{9}$ Muon
  Decays?},'' \href{http://dx.doi.org/10.1016/0370-2693(77)90435-X}{{ Phys.
  Lett. B} {\bfseries 67} (1977) 421--428}.

\bibitem{Yanagida:1979as}
T.~Yanagida, ``{Horizontal Gauge Symmetry and Masses of Neutrinos},''. In
  Proceedings of the Workshop on the Baryon Number of the Universe and Unified
  Theories, Tsukuba, Japan, p95 (1979).

\bibitem{Corbelli:1999af}
E.~Corbelli and P.~Salucci, ``{The Extended Rotation Curve and the Dark Matter
  Halo of M33},'' \href{http://dx.doi.org/10.1046/j.1365-8711.2000.03075.x}{{
  Mon. Not. Roy. Astron. Soc.} {\bfseries 311} (2000) 441--447},
  \href{http://arxiv.org/abs/astro-ph/9909252}{{\ttfamily
  arXiv:astro-ph/9909252}}.

\bibitem{Sofue:2000jx}
Y.~Sofue and V.~Rubin, ``{Rotation Curves of Spiral Galaxies},''
  \href{http://dx.doi.org/10.1146/annurev.astro.39.1.137}{{ Ann. Rev. Astron.
  Astrophys.} {\bfseries 39} (2001) 137--174},
  \href{http://arxiv.org/abs/astro-ph/0010594}{{\ttfamily
  arXiv:astro-ph/0010594}}.

\bibitem{Massey:2010hh}
R.~Massey, T.~Kitching, and J.~Richard, ``{The Dark Matter of Gravitational
  Lensing},'' \href{http://dx.doi.org/10.1088/0034-4885/73/8/086901}{{ Rept.
  Prog. Phys.} {\bfseries 73} (2010) 086901},
  \href{http://arxiv.org/abs/1001.1739}{{\ttfamily arXiv:1001.1739
  [astro-ph.CO]}}.

\bibitem{Aghanim:2018eyx}
{\bfseries Planck} Collaboration, N.~Aghanim {et al.}, ``{Planck 2018 Results.
  VI. Cosmological Parameters},''
  \href{http://dx.doi.org/10.1051/0004-6361/201833910}{{ Astron. Astrophys.}
  {\bfseries 641} (2020) A6}, \href{http://arxiv.org/abs/1807.06209}{{\ttfamily
  arXiv:1807.06209 [astro-ph.CO]}}.

\bibitem{Randall:2007ph}
S.~W. Randall, M.~Markevitch, D.~Clowe, A.~H. Gonzalez, and M.~Bradac,
  ``{Constraints on the Self-Interaction Cross-Section of Dark Matter from
  Numerical Simulations of the Merging Galaxy Cluster 1E 0657-56},''
  \href{http://dx.doi.org/10.1086/587859}{{ Astrophys. J.} {\bfseries 679}
  (2008) 1173--1180}, \href{http://arxiv.org/abs/0704.0261}{{\ttfamily
  arXiv:0704.0261 [astro-ph]}}.

\bibitem{Arcadi:2017kky}
G.~Arcadi, M.~Dutra, P.~Ghosh, M.~Lindner, Y.~Mambrini, M.~Pierre, S.~Profumo,
  and F.~S. Queiroz, ``{The Waning of the WIMP? A Review of Models, Searches,
  and Constraints},'' \href{http://dx.doi.org/10.1140/epjc/s10052-018-5662-y}{{
  Eur. Phys. J. C} {\bfseries 78} no.~3, (2018) 203},
  \href{http://arxiv.org/abs/1703.07364}{{\ttfamily arXiv:1703.07364
  [hep-ph]}}.

\bibitem{Freytsis:2010ne}
M.~Freytsis and Z.~Ligeti, ``{On Dark Matter Models with Uniquely
  Spin-Dependent Detection Possibilities},''
  \href{http://dx.doi.org/10.1103/PhysRevD.83.115009}{{ Phys. Rev. D}
  {\bfseries 83} (2011) 115009},
  \href{http://arxiv.org/abs/1012.5317}{{\ttfamily arXiv:1012.5317 [hep-ph]}}.

\bibitem{Ipek:2014gua}
S.~Ipek, D.~McKeen, and A.~E. Nelson, ``{A Renormalizable Model for the
  Galactic Center Gamma Ray Excess from Dark Matter Annihilation},''
  \href{http://dx.doi.org/10.1103/PhysRevD.90.055021}{{ Phys. Rev. D}
  {\bfseries 90} no.~5, (2014) 055021},
  \href{http://arxiv.org/abs/1404.3716}{{\ttfamily arXiv:1404.3716 [hep-ph]}}.

\bibitem{Arcadi:2017wqi}
G.~Arcadi, M.~Lindner, F.~S. Queiroz, W.~Rodejohann, and S.~Vogl,
  ``{Pseudoscalar Mediators: A WIMP Model at the Neutrino Floor},''
  \href{http://dx.doi.org/10.1088/1475-7516/2018/03/042}{{ JCAP} {\bfseries 03}
  (2018) 042}, \href{http://arxiv.org/abs/1711.02110}{{\ttfamily
  arXiv:1711.02110 [hep-ph]}}.

\bibitem{Sanderson:2018lmj}
N.~F. Bell, G.~Busoni, and I.~W. Sanderson, ``{Loop Effects in Direct
  Detection},'' \href{http://dx.doi.org/10.1088/1475-7516/2018/08/017}{{ JCAP}
  {\bfseries 08} (2018) 017}, \href{http://arxiv.org/abs/1803.01574}{{\ttfamily
  arXiv:1803.01574 [hep-ph]}}. [Erratum: JCAP 01, E01 (2019)].

\bibitem{Abe:2018emu}
T.~Abe, M.~Fujiwara, and J.~Hisano, ``{Loop Corrections to Dark Matter Direct
  Detection in a Pseudoscalar Mediator Dark Matter Model},''
  \href{http://dx.doi.org/10.1007/JHEP02(2019)028}{{ JHEP} {\bfseries 02}
  (2019) 028}, \href{http://arxiv.org/abs/1810.01039}{{\ttfamily
  arXiv:1810.01039 [hep-ph]}}.

\bibitem{Abe:2019wjw}
T.~Abe, M.~Fujiwara, J.~Hisano, and Y.~Shoji, ``{Maximum Value of the
  Spin-Independent Cross Section in the 2HDM+a},''
  \href{http://dx.doi.org/10.1007/JHEP01(2020)114}{{ JHEP} {\bfseries 01}
  (2020) 114}, \href{http://arxiv.org/abs/1910.09771}{{\ttfamily
  arXiv:1910.09771 [hep-ph]}}.

\bibitem{Barger:2010yn}
V.~Barger, M.~McCaskey, and G.~Shaughnessy, ``{Complex Scalar Dark Matter
  vis-a-vis CoGeNT, DAMA/LIBRA and XENON100},''
  \href{http://dx.doi.org/10.1103/PhysRevD.82.035019}{{ Phys. Rev. D}
  {\bfseries 82} (2010) 035019},
  \href{http://arxiv.org/abs/1005.3328}{{\ttfamily arXiv:1005.3328 [hep-ph]}}.

\bibitem{Barducci:2016fue}
D.~Barducci, A.~Bharucha, N.~Desai, M.~Frigerio, B.~Fuks, A.~Goudelis,
  S.~Kulkarni, G.~Polesello, and D.~Sengupta, ``{Monojet Searches for
  Momentum-Dependent Dark Matter Interactions},''
  \href{http://dx.doi.org/10.1007/JHEP01(2017)078}{{ JHEP} {\bfseries 01}
  (2017) 078}, \href{http://arxiv.org/abs/1609.07490}{{\ttfamily
  arXiv:1609.07490 [hep-ph]}}.

\bibitem{Gross:2017dan}
C.~Gross, O.~Lebedev, and T.~Toma, ``{Cancellation Mechanism for
  Dark-Matter--Nucleon Interaction},''
  \href{http://dx.doi.org/10.1103/PhysRevLett.119.191801}{{ Phys. Rev. Lett.}
  {\bfseries 119} no.~19, (2017) 191801},
  \href{http://arxiv.org/abs/1708.02253}{{\ttfamily arXiv:1708.02253
  [hep-ph]}}.

\bibitem{Balkin:2017aep}
R.~Balkin, M.~Ruhdorfer, E.~Salvioni, and A.~Weiler, ``{Charged Composite
  Scalar Dark Matter},'' \href{http://dx.doi.org/10.1007/JHEP11(2017)094}{{
  JHEP} {\bfseries 11} (2017) 094},
  \href{http://arxiv.org/abs/1707.07685}{{\ttfamily arXiv:1707.07685
  [hep-ph]}}.

\bibitem{Ishiwata:2018sdi}
K.~Ishiwata and T.~Toma, ``{Probing Pseudo Nambu-Goldstone Boson Dark Matter at
  Loop Level},'' \href{http://dx.doi.org/10.1007/JHEP12(2018)089}{{ JHEP}
  {\bfseries 12} (2018) 089}, \href{http://arxiv.org/abs/1810.08139}{{\ttfamily
  arXiv:1810.08139 [hep-ph]}}.

\bibitem{Huitu:2018gbc}
K.~Huitu, N.~Koivunen, O.~Lebedev, S.~Mondal, and T.~Toma, ``{Probing
  Pseudo-Goldstone Dark Matter at the LHC},''
  \href{http://dx.doi.org/10.1103/PhysRevD.100.015009}{{ Phys. Rev. D}
  {\bfseries 100} no.~1, (2019) 015009},
  \href{http://arxiv.org/abs/1812.05952}{{\ttfamily arXiv:1812.05952
  [hep-ph]}}.

\bibitem{Cline:2019okt}
J.~M. Cline and T.~Toma, ``{Pseudo-Goldstone Dark Matter Confronts Cosmic Ray
  and Collider Anomalies},''
  \href{http://dx.doi.org/10.1103/PhysRevD.100.035023}{{ Phys. Rev. D}
  {\bfseries 100} no.~3, (2019) 035023},
  \href{http://arxiv.org/abs/1906.02175}{{\ttfamily arXiv:1906.02175
  [hep-ph]}}.

\bibitem{Jiang:2019soj}
X.-M. Jiang, C.~Cai, Z.-H. Yu, Y.-P. Zeng, and H.-H. Zhang,
  ``{Pseudo-Nambu-Goldstone Dark Matter and Two-Higgs-Doublet Models},''
  \href{http://dx.doi.org/10.1103/PhysRevD.100.075011}{{ Phys. Rev. D}
  {\bfseries 100} no.~7, (2019) 075011},
  \href{http://arxiv.org/abs/1907.09684}{{\ttfamily arXiv:1907.09684
  [hep-ph]}}.

\bibitem{Arina:2019tib}
C.~Arina, A.~Beniwal, C.~Degrande, J.~Heisig, and A.~Scaffidi, ``{Global Fit of
  Pseudo-Nambu-Goldstone Dark Matter},''
  \href{http://dx.doi.org/10.1007/JHEP04(2020)015}{{ JHEP} {\bfseries 04}
  (2020) 015}, \href{http://arxiv.org/abs/1912.04008}{{\ttfamily
  arXiv:1912.04008 [hep-ph]}}.

\bibitem{Karamitros:2019ewv}
D.~Karamitros, ``{Pseudo Nambu-Goldstone Dark Matter: Examples of Vanishing
  Direct Detection Cross Section},''
  \href{http://dx.doi.org/10.1103/PhysRevD.99.095036}{{ Phys. Rev. D}
  {\bfseries 99} no.~9, (2019) 095036},
  \href{http://arxiv.org/abs/1901.09751}{{\ttfamily arXiv:1901.09751
  [hep-ph]}}.

\bibitem{Abe:2020iph}
Y.~Abe, T.~Toma, and K.~Tsumura, ``{Pseudo-Nambu-Goldstone Dark Matter from
  Gauged $U(1)_{B-L}$ Symmetry},''
  \href{http://dx.doi.org/10.1007/JHEP05(2020)057}{{ JHEP} {\bfseries 05}
  (2020) 057}, \href{http://arxiv.org/abs/2001.03954}{{\ttfamily
  arXiv:2001.03954 [hep-ph]}}.

\bibitem{Okada:2020zxo}
N.~Okada, D.~Raut, and Q.~Shafi, ``{Pseudo-Goldstone Dark Matter in a Gauged
  $B-L$ Extended Standard Model},''
  \href{http://dx.doi.org/10.1103/PhysRevD.103.055024}{{ Phys. Rev. D}
  {\bfseries 103} no.~5, (2021) 055024},
  \href{http://arxiv.org/abs/2001.05910}{{\ttfamily arXiv:2001.05910
  [hep-ph]}}.

\bibitem{Zhang:2021alu}
Z.~Zhang, C.~Cai, X.-M. Jiang, Y.-L. Tang, Z.-H. Yu, and H.-H. Zhang, ``{Phase
  Transition Gravitational Waves from Pseudo-Nambu-Goldstone Dark Matter and
  Two Higgs Doublets},'' \href{http://dx.doi.org/10.1007/JHEP05(2021)160}{{
  JHEP} {\bfseries 05} (2021) 160},
  \href{http://arxiv.org/abs/2102.01588}{{\ttfamily arXiv:2102.01588
  [hep-ph]}}.

\bibitem{Abe:2021jcz}
T.~Abe, ``{Early kinetic decoupling and a pseudo-Nambu-Goldstone dark matter
  model},'' \href{http://dx.doi.org/10.1103/PhysRevD.104.035025}{{ Phys. Rev.
  D} {\bfseries 104} no.~3, (2021) 035025},
  \href{http://arxiv.org/abs/2106.01956}{{\ttfamily arXiv:2106.01956
  [hep-ph]}}.

\bibitem{Abe:2021vat}
Y.~Abe and T.~Toma, ``{Direct Detection of Pseudo-Nambu-Goldstone Dark Matter
  with Light Mediator},''
  \href{http://dx.doi.org/10.1016/j.physletb.2021.136639}{{ Phys. Lett. B}
  {\bfseries 822} (2021) 136639},
  \href{http://arxiv.org/abs/2108.10647}{{\ttfamily arXiv:2108.10647
  [hep-ph]}}.

\bibitem{Zeng:2021moz}
Y.-P. Zeng, X.~Xiao, and W.~Wang, ``{Constraints on Pseudo-Nambu-Goldstone Dark
  Matter from Direct Detection Experiment and Neutron Star Reheating
  Temperature},'' \href{http://dx.doi.org/10.1016/j.physletb.2021.136822}{{
  Phys. Lett. B} {\bfseries 824} (2022) 136822},
  \href{http://arxiv.org/abs/2108.11381}{{\ttfamily arXiv:2108.11381
  [hep-ph]}}.

\bibitem{Abe:2022mlc}
T.~Abe and Y.~Hamada, ``{A Model of Pseudo-Nambu\textendash{}Goldstone Dark
  Matter from a Softly Broken $SU(2)$ Global Symmetry with a $U(1)$ Gauge
  Symmetry},'' \href{http://dx.doi.org/10.1093/ptep/ptad021}{{ PTEP} {\bfseries
  2023} no.~3, (2023) 033B04},
  \href{http://arxiv.org/abs/2205.11919}{{\ttfamily arXiv:2205.11919
  [hep-ph]}}.

\bibitem{Otsuka:2022zdy}
H.~Otsuka, T.~Shimomura, K.~Tsumura, Y.~Uchida, and N.~Yamatsu,
  ``{Pseudo-Nambu-Goldstone Dark Matter from Non-Abelian Gauge Symmetry},''
  \href{http://dx.doi.org/10.1103/PhysRevD.106.115033}{{ Phys. Rev. D}
  {\bfseries 106} no.~11, (2022) 115033},
  \href{http://arxiv.org/abs/2210.08696}{{\ttfamily arXiv:2210.08696
  [hep-ph]}}.

\bibitem{Abe:2021byq}
Y.~Abe, T.~Toma, K.~Tsumura, and N.~Yamatsu, ``{Pseudo-Nambu-Goldstone Dark
  Matter Model Inspired by Grand Unification},''
  \href{http://dx.doi.org/10.1103/PhysRevD.104.035011}{{ Phys. Rev. D}
  {\bfseries 104} (2021) 035011},
  \href{http://arxiv.org/abs/2104.13523}{{\ttfamily arXiv:2104.13523
  [hep-ph]}}.

\bibitem{Okada:2021qmi}
N.~Okada, D.~Raut, Q.~Shafi, and A.~Thapa, ``{Pseudo-Goldstone dark matter in
  $SO(10)$},'' \href{http://dx.doi.org/10.1103/PhysRevD.104.095002}{{ Phys.
  Rev. D} {\bfseries 104} no.~9, (2021) 095002},
  \href{http://arxiv.org/abs/2105.03419}{{\ttfamily arXiv:2105.03419
  [hep-ph]}}.

\bibitem{Feger:2019tvk}
R.~Feger, T.~W. Kephart, and R.~J. Saskowski, ``{LieART 2.0 -- A Mathematica
  Application for Lie Algebras and Representation Theory},'' { Comput. Phys.
  Commun.} {\bfseries 257} (2020) 107490,
\href{http://arxiv.org/abs/1912.10969}{{\ttfamily arXiv:1912.10969 [hep-th]}}.

\bibitem{Fonseca:2020vke}
R.~M. Fonseca, ``{GroupMath: A Mathematica Package for Group Theory
  Calculations},'' \href{http://dx.doi.org/10.1016/j.cpc.2021.108085}{{ Comput.
  Phys. Commun.} {\bfseries 267} (2021) 108085},
  \href{http://arxiv.org/abs/2011.01764}{{\ttfamily arXiv:2011.01764
  [hep-th]}}.

\bibitem{Fonseca:2017lem}
R.~M. Fonseca, ``{The Sym2Int Program: Going from Symmetries to
  Interactions},'' \href{http://dx.doi.org/10.1088/1742-6596/873/1/012045}{{ J.
  Phys. Conf. Ser.} {\bfseries 873} no.~1, (2017) 012045},
  \href{http://arxiv.org/abs/1703.05221}{{\ttfamily arXiv:1703.05221
  [hep-ph]}}.

\bibitem{Fonseca:2019yya}
R.~M. Fonseca, ``{Enumerating the Operators of an Effective Field Theory},''
  \href{http://dx.doi.org/10.1103/PhysRevD.101.035040}{{ Phys. Rev. D}
  {\bfseries 101} no.~3, (2020) 035040},
  \href{http://arxiv.org/abs/1907.12584}{{\ttfamily arXiv:1907.12584
  [hep-ph]}}.

\bibitem{Foot:1988aq}
R.~Foot, H.~Lew, X.~G. He, and G.~C. Joshi, ``{Seesaw Neutrino Masses Induced
  by a Triplet of Leptons},''
\href{http://dx.doi.org/10.1007/BF01415558}{{ Z. Phys.} {\bfseries C44} (1989)
  441}.

\bibitem{ParticleDataGroup:2022pth}
{\bfseries Particle Data Group} Collaboration, R.~L. Workman {et al.},
  ``{Review of Particle Physics},''
  \href{http://dx.doi.org/10.1093/ptep/ptac097}{{ PTEP} {\bfseries 2022} (2022)
  083C01}.

\bibitem{McKay:1981}
W.~G. McKay and J.~Patera, { Tables of Dimensions, Indices, and Branching Rules
  for Representations of Simple Lie Algebras}.
\newblock Marcel Dekker, Inc., New York, 1981.

\bibitem{Fonseca:2011sy}
R.~M. Fonseca, ``{Calculating the Renormalisation Group Equations of a SUSY
  Model with Susyno},'' \href{http://dx.doi.org/10.1016/j.cpc.2012.05.017}{{
  Comput.Phys.Commun.} {\bfseries 183} (2012) 2298--2306},
\href{http://arxiv.org/abs/1106.5016}{{\ttfamily arXiv:1106.5016 [hep-ph]}}.

\bibitem{Feger:2012bs}
R.~Feger and T.~W. Kephart, ``{LieART - A Mathematica Application for Lie
  Algebras and Representation Theory},''
  \href{http://dx.doi.org/10.1016/j.cpc.2014.12.023}{{ Comput.Phys.Commun.}
  {\bfseries 192} (2015) 166--195},
\href{http://arxiv.org/abs/1206.6379}{{\ttfamily arXiv:1206.6379 [math-ph]}}.

\bibitem{Mohapatra2002}
R.~N. Mohapatra, { {Unification and Supersymmetry -The Frontiers of
  Quarks-Lepton Physics-}}.
\newblock Springer, 2002.

\bibitem{Nath:2006ut}
P.~Nath and P.~Fileviez~Perez, ``{Proton Stability in Grand Unified Theories,
  in Strings and in Branes},''
  \href{http://dx.doi.org/10.1016/j.physrep.2007.02.010}{{ Phys. Rept.}
  {\bfseries 441} (2007) 191--317},
  \href{http://arxiv.org/abs/hep-ph/0601023}{{\ttfamily arXiv:hep-ph/0601023}}.

\bibitem{LZ:2022lsv}
{\bfseries LZ} Collaboration, J.~Aalbers {et al.}, ``{First Dark Matter Search
  Results from the LUX-ZEPLIN (LZ) Experiment},''
  \href{http://dx.doi.org/10.1103/PhysRevLett.131.041002}{{ Phys. Rev. Lett.}
  {\bfseries 131} no.~4, (2023) 041002},
  \href{http://arxiv.org/abs/2207.03764}{{\ttfamily arXiv:2207.03764
  [hep-ex]}}.

\bibitem{ATLAS:2020kdi}
{\bfseries ATLAS} Collaboration, T.~A. Collaboration, ``{Combination of
  Searches for Invisible Higgs Boson Decays with the ATLAS Experiment},'' {
  ATLAS-CONF-2020-052} (2020) .

\bibitem{FCC:2018vvp}
{\bfseries FCC} Collaboration, A.~Abada {et al.}, ``{FCC-hh: The Hadron
  Collider}: {Future Circular Collider Conceptual Design Report Volume 3},''
  \href{http://dx.doi.org/10.1140/epjst/e2019-900087-0}{{ Eur. Phys. J. ST}
  {\bfseries 228} no.~4, (2019) 755--1107}.

\bibitem{Delahaye:2019omf}
J.~P. Delahaye, M.~Diemoz, K.~Long, B.~Mansouli\'e, N.~Pastrone, L.~Rivkin,
  D.~Schulte, A.~Skrinsky, and A.~Wulzer, ``{Muon Colliders},''
  \href{http://arxiv.org/abs/1901.06150}{{\ttfamily arXiv:1901.06150
  [physics.acc-ph]}}.

\end{thebibliography}\endgroup

\end{document}